\documentclass[12pt, draftclsnofoot, onecolumn]{IEEEtran}
\ifCLASSINFOpdf
\else
\fi
\usepackage[cmex10]{amsmath}
\usepackage{upgreek}
\usepackage{booktabs}
\usepackage{epsfig}
\usepackage{latexsym}
\usepackage{multirow}
\usepackage{stfloats}
\usepackage{epstopdf}
\usepackage{color}  
\usepackage{tabularx} 
\usepackage{amssymb}
\usepackage{enumerate}
\graphicspath{{./Figures/}}
\usepackage{color}
\usepackage{bbm}
\usepackage{bm}
\usepackage{cite}
\usepackage[tight,footnotesize]{subfigure}
\usepackage{balance}
\usepackage{mathrsfs}
\usepackage{verbatim}
\usepackage{dsfont}
\usepackage{verbatim}
\usepackage{tikz}
\usepackage{diagbox}
\usepackage{caption}
\usepackage[framemethod=tikz]{mdframed}
\usepackage{multicol}
\usepackage{environ}
\usepackage{tikz}
\usepackage{amsmath}
\usepackage{stfloats}

\begin{document}
\title{Joint Inter-path and Intra-path Multiplexing for Terahertz Widely-spaced Multi-subarray Hybrid Beamforming Systems}
\author{
Longfei Yan, Yuhang Chen,
Chong Han,~\IEEEmembership{Member,~IEEE},
and \\Jinhong Yuan,~\IEEEmembership{Fellow,~IEEE}
\thanks{
This paper was presented in part at the IEEE INFOCOM workshops, Paris, France, April 29-May 2, 2019~\cite{11}.

Longfei Yan, Yuhang Chen, and Chong Han are with the Terahertz Wireless Communications (TWC) Laboratory, Shanghai Jiao Tong University, Shanghai 200240, China (e-mail: \{longfei.yan, yuhang.chen, chong.han\}@sjtu.edu.cn).

Jinhong Yuan is with the School of Electrical Engineering and Telecommunications, University of New South Wales, Sydney, NSW 2052, Australia (e-mail: j.yuan@unsw.edu.au).}}
\maketitle	
\begin{abstract}
Terahertz (THz) communications with multi-GHz bandwidth are envisioned as a key technology for 6G systems. Ultra-massive (UM) MIMO with hybrid beamforming architectures are widely investigated to provide a high array gain to overcome the huge propagation loss. However, most of the existing hybrid beamforming architectures can only utilize the multiplexing offered by the multipath components, i.e., inter-path multiplexing, which is very limited due to the spatially sparse THz channel. In this paper, a widely-spaced multi-subarray (WSMS) hybrid beamforming architecture is proposed, which improves the multiplexing gain by exploiting a new type of intra-path multiplexing provided by the spherical-wave propagation among $k$ widely-spaced subarrays, in addition to the inter-path multiplexing. The resulting multiplexing gain of WSMS architecture is $k$ times of the existing architectures. To harness WSMS hybrid beamforming, a novel design problem is formulated by optimizing the number of subarrays, subarray spacing, and hybrid beamforming matrices to maximize the spectral efficiency, which is decomposed into two subproblems. An optimal closed-form solution is derived for the first hybrid beamforming subproblem, while a dominant-line-of-sight-relaxation algorithm is proposed for the second array configuration subproblem. Extensive simulation results demonstrate that the WSMS architecture and proposed algorithms substantially enhance the spectral efficiency and energy efficiency.
\end{abstract}
\IEEEpeerreviewmaketitle
\section{Introduction}
\IEEEPARstart{W}{ith} unprecedented multi-GHz bandwidth, the Terahertz (THz) band has drawn increasing attention to support 100+ Gbps wireless data rates~\cite{8732419}.
Recently, the Federal Communications Commission (FCC) created a new category of experimental licenses for the use of frequencies between 95 GHz and 3 THz, as preparation for 6G THz wireless systems.
Although with ultra-broad bandwidth, the THz band suffers from huge propagation loss, which significantly limits the wireless communication distance~\cite{8387211}.
Thanks to the sub-millimeter wavelength, design of array consisting of 512 and even 1024 antennas at transceivers is feasible, which enables THz ultra-massive multiple-input multiple-output (UM-MIMO) systems~\cite{AKYILDIZ201646,sarieddeen2019terahertzband}. The multi-antenna system can generate a high array gain to compensate the path loss and solve the distance problem, while offer a multiplexing gain to further improve the spectral efficiency of the THz communications. 

In the THz band, many hardware constraints preclude from using conventional digital beamforming, which, instead, motivates the appealing hybrid beamforming technology~\cite{7786122,8531683,7914742,8651536}.
The hybrid beamforming divides the signal processing into the digital baseband domain and analog RF domain, which can achieve high spectral efficiency while maintaining a reasonably low hardware complexity~\cite{7786122,1}. The fully-connected (FC) and array-of-subarrays (AoSA) architectures are two classic hybrid beamforming architectures~\cite{1,7397861,7389996,7445130}. In the FC architecture, each RF chain connects to all antennas. As a critical difference, each RF chain only connects to a subset of antennas, i.e., a subarray, in the AoSA architecture. Furthermore, the dynamic architectures which own adaptive connections between RF chains and antennas to achieve more flexible trade-off between the spectral efficiency and power consumption are investigated~\cite{DAoSA_JSAC_2020,7880698,9110865}. 

Most of the existing hybrid beamforming architectures use the antenna array with $0.5\lambda$ antenna spacing, where $\lambda$ denotes the wavelength~\cite{1,7397861,DAoSA_JSAC_2020,7445130,7389996}. Based on the half-wavelength antenna array, the planar-wave assumption is considered and the spatial multiplexing only benefits from exploiting different multipath components, which is referred to \textit{inter-path multiplexing}~\cite{11,8356240}. The inter-path multiplexing gain is upper-bounded by the number of resolvable multipath.
Due to the prohibitively high propagation attenuation and scattering loss, the THz UM-MIMO channel is usually spatially sparse and the number of multipath is very limited, e.g., typically around 5~\cite{6998944,7786122,DAoSA_JSAC_2020}. As a result, even with high array gain to enhance the received power, the poor inter-path multiplexing gain at THz band still significantly restricts the spectral efficiency of the existing hybrid beamforming architectures. Consequently, it is highly desirable to improve the multiplexing gain of THz hybrid beamforming architectures via other means.

An interesting approach to the above problem is harvesting a new type of multiplexing, i.e., \textit{intra-path multiplexing}~\cite{Song2017,8466787,4155681,7501567}. The key idea of intra-path multiplexing is enlarging the antenna spacing to make the planar-wave assumption invalid, and instead, considering the spherical-wave propagation among antennas. Under spherical-wave propagation, the phases of one propagation path on different antennas become linearly independent. 
As a result, multiplexing gain can be exploited for each propagation path but among different antennas, namely, intra-path multiplexing. 
The intra-path multiplexing has been investigated in line-of-sight (LoS) MIMO architecture at microwave and mmWave frequencies to offer multiplexing gain with only one LoS path~\cite{Song2017,8466787,4155681,7501567}.
Since the intra-path multiplexing gain is not limited by the number of multipath, it is naturally applicable and promising for THz communications with sparse channel.

\begin{figure*}
	\centering
	\captionsetup{font={footnotesize}}
	\includegraphics[scale=0.4]{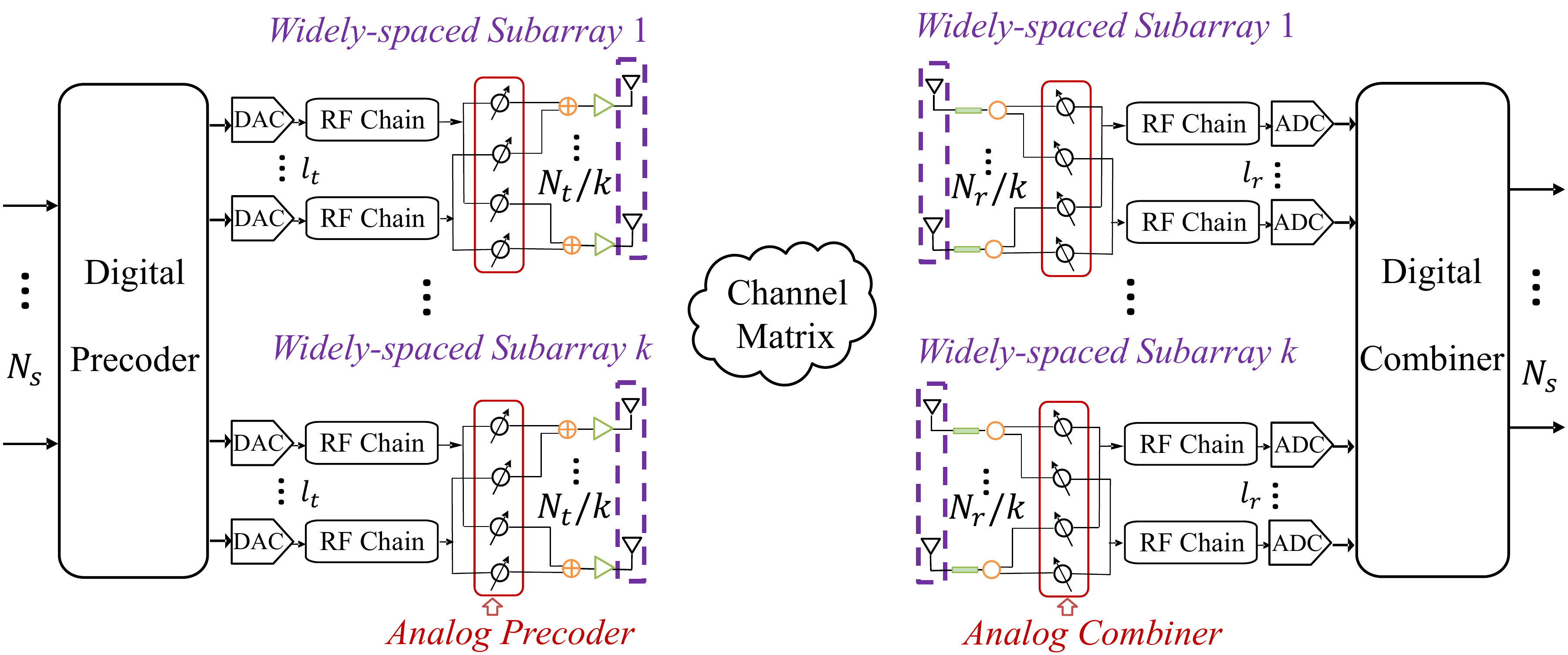}
	\caption{The architecture of the THz WSMS hybrid beamforming. $l_t$ and $l_r$ are the number of RF chains connected with one subarray. $N_s$ is the number of data streams.}
	\label{architecture}
	\vspace{-8.5mm}
\end{figure*}

To harvest the abundant intra-path multiplexing in addition to the poor inter-path multiplexing in THz hybrid beamforming systems, a widely-spaced multi-subarray (WSMS) hybrid beamforming architecture is proposed~\cite{11,8356240}, as shown in Fig.~\ref{architecture}, where $N_t$ and $N_r$ denote the numbers of antennas at transmitter and receiver, respectively. 
The antennas array is uniformly divided into $k$ subarrays. Here the antenna spacing within each subarray is $0.5\lambda$, while the subarrays are widely-spaced to make the array aperture $S$ of whole array exceed $\sqrt{\frac{\lambda D}{2}}$, where $D$ denotes the communication distance. As a result, the Rayleigh distance $\frac{2S^2}{\lambda}$ is larger than $D$, which suggests that the spherical-wave propagation needs to be considered in this near-field propagation.
On one hand, the inter-path multiplexing is provided by the multipath components of THz channel. 
On the other hand, due to the spherical-wave propagation among the widely-spaced subarrays, the phases of each propagation path on different subarrays are linearly independent, which support the intra-path multiplexing.
Compared to the existing hybrid beamforming architectures, additional intra-path multiplexing gain of WSMS architecture can improve the spectral efficiency of THz systems substantially.

Although the idea of WSMS architecture was studied in~\cite{11,8356240},
these studies only investigate a special case of WSMS architecture, i.e., WSMS with two widely-spaced subarrays. However, for a given number of antennas, the number of subarrays $k$ and the subarray spacing $d_s$ make significant impact on the spectral efficiency and energy efficiency of the system. Thus, the careful design of $k$ and $d_s$ of WSMS architecture is critical for the THz systems, which has not been studied yet in~\cite{11,8356240}. Moreover, the WSMS hybrid beamforming algorithms proposed in~\cite{11,8356240} bear high computational complexities. Developing low-complexity yet practical hybrid beamforming algorithms is of significant importance for THz systems.

In this work, we investigate the holistic harvesting of the joint inter-path and intra-path multiplexing for the THz WSMS architecture with general cases that $k\geq2$. We prove that, through the joint utilization of the inter-path and intra-path multiplexing, the multiplexing gain of THz WSMS architecture is $kN_p$, which is $k$ times of that of the existing hybrid beamforming architectures, where $N_p$ denotes the number of multipath.
Furthermore, we analyze that the spectral efficiency of THz WSMS architecture highly depends on $d_s$ and $k$. Inspired by this, we design a novel framework, aiming to jointly optimize $d_s$, $k$, and the hybrid beamforming matrices to maximize the spectral efficiency of the THz systems. Compared to our prior and initial work in~\cite{11} which only considers a special case of WSMS that $k=2$, both the design problem as well as the proposed algorithms in this work are novel and more sophisticated. Moreover, substantially more results and insights for THz system designs are provided in this work.
The distinctive contributions of this work are summarized as follows.
\begin{itemize}
	\item
	\textbf{We derive and analyze the multiplexing gain of THz WSMS architecture.} We first investigate the channel model of the THz WSMS architecture. Then, we prove that the joint inter-path and intra-path multiplexing gain in the THz WSMS architecture is ${kN_p}$, which is $k$ times of the multiplexing gain of the existing THz hybrid beamforming architectures with only inter-path multiplexing. Furthermore, we elaborate that both $d_s$ and $k$ make significant impact on the spectral efficiency of the THz WSMS architecture.
	\item
	\textbf{Inspired by the impact of $d_s$ and $k$, we design a novel framework to harness hybrid beamforming of the THz WSMS architecture,} which aims to jointly design $d_s$, $k$, and the hybrid beamforming matrices to maximize the spectral efficiency. We decompose this intractable design problem into two subproblems. The first subproblem P1 is a hybrid beamforming problem, i.e., designing the hybrid beamforming matrices to maximize the spectral efficiency, by assuming that $d_s$ and $k$ are given and known. The second subproblem P2 is an array configuration problem, i.e., determining $d_s$ and $k$ to maximize the spectral efficiency.
	\item
	\textbf{We propose an optimal closed-form solution for P1 and a dominant-LoS-relaxation (DLR) algorithm for P2.}
	One of the main difficulties of solving hybrid beamforming problem P1 is the block-diagonal constraint caused by the widely-spaced subarrays of WSMS architecture. We propose an optimal closed-form solution to solve the hybrid beamforming problem, by exploiting the structure of the channel of THz WSMS architecture. To solve the array configuration problem P2 with a low-complexity, we propose a DLR algorithm, which transforms the intractable P2 into a tractable form through leveraging the dominant LoS peculiarity of the THz band.
	\item
	\textbf{We carry out extensive simulations to validate the performance of the THz WSMS architecture and the proposed algorithms.} Particularly, using the proposed algorithms, the spectral efficiency of the THz WSMS architecture is 96\% higher than the existing architectures, thanks to the joint utilization of inter-path and intra-path multiplexing gain. Moreover, the energy efficiency is 91\% higher than the existing architectures. Furthermore, we demonstrate that the performances of the proposed optimal closed-form solution and the DLR algorithm are optimal and near-optimal, respectively, with remarkably reduced computational complexities.
\end{itemize} 
The remainder of this paper is organized as follows. Channel model and the analysis of joint inter-path and intra-path multiplexing are presented in Sec.~\ref{section_channel_CS_WS}. We investigate the system model and formulate the design problem for THz WSMS hybrid beamforming architecture in Sec.~\ref{section_system_and_design}. The optimal closed-form hybrid beamforming solution is proposed in Sec.~\ref{section_beamforming} and the DLR algorithm is proposed in Sec.~\ref{section_design_subarray}. Furthermore, extensive simulation results are discussed in Sec. \ref{section_simulation}. Finally, the conclusion is drawn in Sec. \ref{section_conclusion}.
	
\textit{Notations}: $\textbf{A}$ is a matrix; $\textbf{a}$ is a vector; $a$ is a scalar; $\textbf{A}[i,l]$ is the element of the $i^{\mathrm{th}}$ row and $l^{\mathrm{th}}$ column of $\textbf{A}$; $\textbf{a}[i]$ denotes the $i^{\rm th}$ element of $\textbf{a}$; $\textbf{I}_{N}$ is an $N$-dimensional identity matrix; $(\cdot)^T$, $(\cdot)^*$, and $(\cdot)^{H}$ represent transpose, conjugate, and conjugate transpose; $\lVert\cdot\rVert_{F}$ is the Frobenius norm of the matrix; $\lVert\cdot\rVert_{p}$ is the $p$-norm of the vector.

\section{Joint Inter-path and Intra-path Multiplexing in THz WSMS Architecture}
\label{section_channel_CS_WS}
In this section, we first study the existing hybrid beamforming architectures which only utilize the inter-path multiplexing. Then, we investigate the new type of intra-path multiplexing in LoS MIMO architecture, which is not limited by the number of multipath and is uniquely promising for the THz sparse channel. In light of this, we analyze a novel THz WSMS hybrid beamforming architecture which can jointly harvest the inter-path and intra-path multiplexing gain. Furthermore, we prove that the joint inter-path and intra-path multiplexing gain in THz WSMS architecture is $kN_p$, which is $k$ times of the multiplexing gain of the existing hybrid beamforming architectures.

\subsection{Inter-path Multiplexing in Existing Hybrid beamforming Architectures}
\subsubsection{\textbf{Channel model}}
\begin{figure}
	\centering
	\captionsetup{font={footnotesize}}
	\includegraphics[scale=0.37]{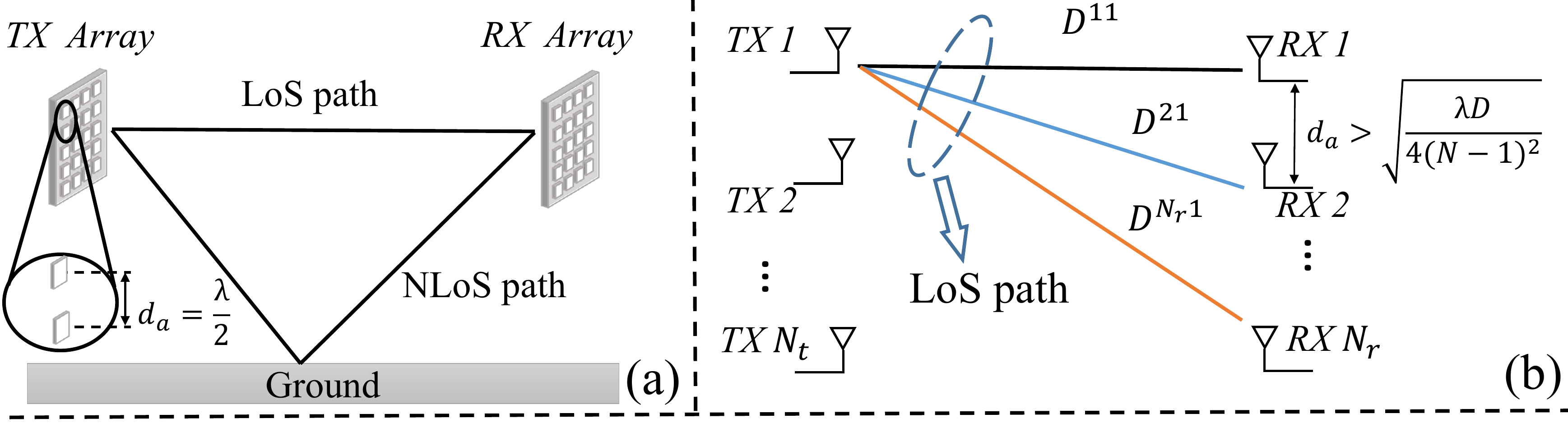}
	\caption{(a) The antenna array of existing hybrid beamforming architectures with $d_a=\frac{\lambda}{2}$. (b) The antenna array of LoS MIMO architecture with widely-spaced antennas.}
	\label{Figure_channel_two}
	\vspace{-6.5mm}
\end{figure}

Most of the existing hybrid beamforming architectures, e.g., FC and AoSA, consider that the antenna spacing $d_{a}$ equals to $0.5\lambda$, as shown in Fig. \ref{Figure_channel_two}(a)~\cite{1,7397861,DAoSA_JSAC_2020,7445130,7389996}. Multiple propagation paths are considered while we only draw two paths for simplicity.
Since the antenna spacing is $0.5\lambda$ which is on the sub-millimeter level at THz band, the array aperture is far less than the communication distance $D$, for which the planar-wave propagation assumption is appropriate. We use $N_t$ and $N_r$ to denote the number of antennas at the transmitter and receiver, respectively.
By considering $N_p$ propagation paths, $N_r\times N_t$-dimensional channel matrix $\textbf{H}_{\rm inter}$ of the existing hybrid beamforming architectures can be stated as~\cite{1,DAoSA_JSAC_2020}
\begin{equation}
\textbf{H}_{\rm inter}=\sum\nolimits_{i=1}^{N_p}\widetilde{\alpha}_{i}\textbf{a}_{ri}(\phi_{ri},\theta_{ri})\textbf{a}_{ti}(\phi_{ti},\theta_{ti})^{H},
\label{channel_planar}
\end{equation}
where $\widetilde{\alpha}_{i}$ is the complex path gain of the $ i^{\mathrm{th}} $ multipath~\cite{6998944}. The vectors $\textbf{a}_{ri}(\phi_{ri},\theta_{ri})$ and $\textbf{a}_{ti}(\phi_{ti},\theta_{ti})$ are the received and transmitted array response vectors of the $i^{\rm th}$ multipath. For an $N_L\times N_W$-element uniform planar array on the x-z plane, $\textbf{a}_{ti}(\phi_{ti},\theta_{ti})$ can be written as~\cite{8493600}
\begin{equation}
\begin{aligned}
\textbf{a}_{t_i}(\phi_{ti},&\theta_{ti})\!\!=\!\!\big[1,..., e^{j\frac{2\pi}{\lambda}d_{a}(n_L\cdot{\rm sin}(\theta_{ti}){\rm cos}(\phi_{ti})+n_W\cdot{\rm cos}(\theta_{ti}))},... ,e^{j\frac{2\pi}{\lambda}d_{a}((N_L-1){\rm sin}(\theta_{ti}){\rm cos}(\phi_{ti})+(N_W-1){\rm cos}(\theta_{ti}))}\big]^T\!\!,
\end{aligned}
\label{steering_vector}
\end{equation}
where $\phi_{ti}$ and $\theta_{ti}$ are the azimuth and elevation departure angles of the $i^{\rm th}$ multipath, with $0\leq n_L\leq (N_L-1)$ and $0\leq n_W\leq (N_W-1)$, respectively. For $\textbf{a}_{r_i}(\phi_{ri},\theta_{ri})$, the subscript ${ti}$ in \eqref{steering_vector} should be replaced by ${ri}$.

\subsubsection{\textbf{Inter-path multiplexing}}
The channel $\textbf{H}_{\rm inter}$ can be decomposed as $N_{\rm rank}$ parallel sub-channels via the singular value decomposition (SVD), where $N_{\rm rank}={\rm min}\{N_t,N_r,N_p\}$ is the rank of $\textbf{H}_{\rm inter}$~\cite{1,11,8356240}. In the THz band, due to the prohibitively high reflection and scattering loss, the contribution of the higher-order reflection paths and the scattering paths is negligible. As a result, the number of multipath $N_p$ is low, e.g., typically around 5~\cite{7786122,DAoSA_JSAC_2020}. Hence, $N_{\rm rank}$ usually equals to $N_p$ such that the spatial multiplexing gain is $N_p$. Benefiting from the different multipath, this gain is referred to the inter-path multiplexing gain~\cite{11,8356240}. Since $N_p$ is usually limited at THz band, the inter-path multiplexing gain is very poor. Even with high array gain to enhance the received power, the poor inter-path multiplexing gain still restricts the spectral efficiency significantly. Therefore, finding other ways to enhance the multiplexing gain in THz hybrid beamforming architectures is highly desirable.

\subsection{Intra-path Multiplexing in LoS MIMO Architecture}
There exists intra-path multiplexing offered by spherical-wave propagation, which is not limited by the number of multipath and is promising for THz systems with sparse channel. To analyze the intra-path multiplexing gain, we consider a LoS MIMO system with only LoS path such that there is no inter-path multiplexing, as shown in Fig.~\ref{Figure_channel_two}(b).

\subsubsection{\textbf{Channel model}}
We denote $S$ as the array aperture of the antenna array, which equals to the length of the diagonal of the uniform planar array. Let us consider $N_t=N_r=N^2$ and $S=\sqrt{2}(N-1)d_{a}$. In a LoS MIMO architecture, the antenna spacing $d_{a}$ is set to be larger than $\sqrt{\frac{\lambda D}{4(N-1)^2}}$ to make $S>\sqrt{\frac{\lambda D}{2}}$, where $D$ is the communication distance. As a result, the Rayleigh distance $D_{ray}=\frac{2S^2}{\lambda}$ of the array is larger than $D$, which implies that the planar-wave assumption is invalid and the spherical-wave propagation needs to be considered~\cite{4155681,7501567}.
Hence, the channel between the $n^{\rm th}$ received antenna and the $m^{\rm th}$ transmitted antenna is denoted as $\lvert\widetilde{\alpha}^{mn}\rvert e^{j\frac{2\pi}{\lambda}D^{mn}}$, where $\lvert\widetilde{\alpha}^{mn}\rvert$ denotes the magnitude of the path gain, and $D^{mn}$ represents the distance between these two antennas. 
It has been investigated in~\cite{11,8356240} that although the array aperture $S$ is larger than $\sqrt{\frac{\lambda D}{2}}$, it is still on the order of $\sqrt{\frac{\lambda D}{2}}$ and much smaller than $D$, due to the sub-millimeter wavelength at THz band.
Therefore, all $\lvert\widetilde{\alpha}^{mn}\rvert$ can be treated identical and approximated as $\lvert\widetilde{\alpha}\rvert$. Hence, the $N_r\times N_t$-dimensional channel matrix $\textbf{H}_{\rm intra}$ of the LoS MIMO architecture is described as~\cite{Song2017,8466787,4155681,7501567}
\begin{equation}
\textbf{H}_{\rm intra}=\lvert\widetilde{\alpha}\rvert\textbf{G},
\label{channel model_LOS}
\end{equation} 
where $\textbf{G}$ denotes the phase matrix such that $\textbf{G}[m,n]=e^{j\frac{2\pi}{\lambda}D^{mn}}$.
	
\subsubsection{\textbf{Intra-path multiplexing}} 
Due to the consideration of spherical-wave propagation, the phases in $\textbf{G}$ of LoS path for different antennas are linearly independent such that the rank of $\textbf{H}_{\rm intra}$ equals to ${\rm min}\{N_t,N_r\}$~\cite{Song2017,4155681}, where $N_t$ and $N_r$ are the number of antennas at transmitter and receiver, respectively. As a result, the value of spatial multiplexing gain is equal to ${\rm min}\{N_t,N_r\}$. Distinguished from the inter-path multiplexing gain which is limited by the number of multipath, this intra-path multiplexing gain is provided by the spherical-wave propagation among antennas and is not limited by the number of multipath. 
Consequently, the intra-path multiplexing is very attractive and promising for THz communications with sparse channel. 

\subsection{Joint Inter-path and Intra-path Multiplexing of THz WSMS Hybrid Beamforming Architecture}
In this subsection, we propose to harvest the promising intra-path multiplexing in THz WSMS architecture, in addition to the poor inter-path multiplexing. We first investigate the channel model of THz WSMS architecture. Then, we analyze the value of joint inter-path and intra-path multiplexing gain of the THz WSMS architecture.

\begin{figure}
	\centering
	\captionsetup{font={footnotesize}}
	\includegraphics[scale=0.29]{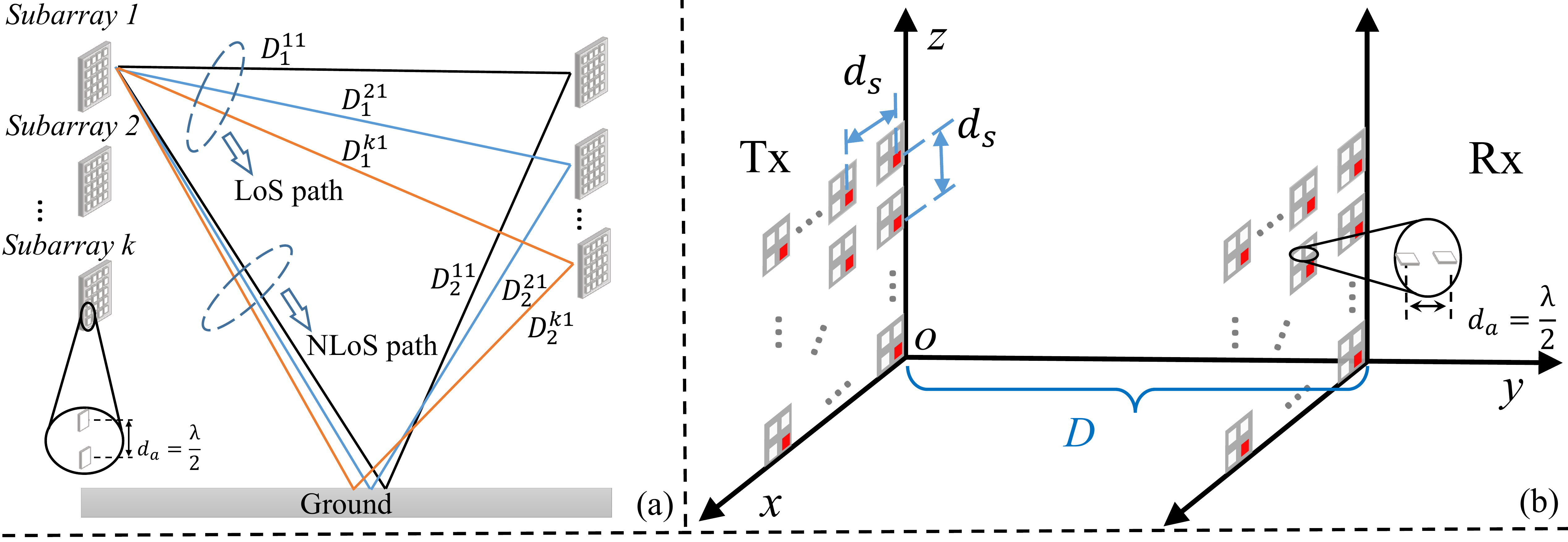}
	\caption{(a) The sketch of the inter-path and intra-path multiplexing in THz WSMS architecture. (b) The antenna array of the THz WSMS architecture.}
	\label{Figure_subarray}
	\vspace{-8.5mm}
\end{figure}

We draw the sketch of the joint utilization of inter-path and intra-path multiplexing in Fig.~\ref{Figure_subarray}(a). The inter-path multiplexing is offered by the multipath components. Inspired by the widely-spaced antennas in LoS MIMO architecture, the array of the WSMS architecture is divided into $k$ widely-spaced subarrays to make the array aperture of the whole array exceed $\sqrt{\frac{\lambda D}{2}}$. As a result, the spherical-wave propagation needs to be considered among the widely-spaced subarrays, which provides the intra-path multiplexing~\cite{11,8356240}. It is worth noting that, thanks to the sub-millimeter wavelength at THz band, the array aperture is still a reasonable value to be implement in practical communication devices. For instance, considering $f=0.3$ THz and $D=40$m, $\sqrt{\frac{\lambda D}{2}}\approx0.14$m. 

\subsubsection{\textbf{Channel model}}
The detailed setting of the antenna array of the THz WSMS architecture is shown in Fig.~\ref{Figure_subarray}(b).
The antennas are uniformly divided into $k$ subarrays, which are represented by the gray patches. 
In each subarray, we use the red and white elements to denote the antennas, where the antenna spacing is $d_a=\frac{\lambda}{2}$. For each subarray, we select its first antenna as the reference point, which is colored in red. We denote the subarray spacing $d_s$ as the distance between the reference points of the adjacent subarrays. $D^{mn}_{i}$ represents the distance between the $n^{\rm th}$ transmitted reference point and the $m^{\rm th}$ received reference point along the $i^{\rm th}$ multipath.  
In this work, we consider that the transmitted and received arrays are on the x-z plane and parallel with each other, as shown in Fig.~\ref{Figure_subarray}(b). The calculation of $D_i^{mn}$ is presented as follows, while for more general cases, the calculation is similar and extensible.
For the $n^{\rm th}$ transmitted reference point, the coordinates are $(x_nd_{s},0,z_nd_{s})$, where $x_n$ and $z_n$ are the indices of reference point. Similarly, for the $m^{\rm th}$ received reference point, the coordinates are $(x_md_{s},D,z_md_{s})$, where $D$ denotes the communication distance. Hence, for LoS path, $D^{mn}_{1}=\sqrt{((x_m-x_n)d_{s})^2+D^2+((z_m-z_n)d_{s})^2}$. For other NLoS paths, $D_{i}^{mn}$ can be calculated with the similar method by considering the locations of reflectors or scatters and the length of the corresponding NLoS paths.

On one hand, the spherical-wave propagation needs to be considered among subarrays, since the whole array aperture is larger than $\sqrt{\frac{\lambda D}{2}}$. Hence, the channel elements among subarrays follow the similar form with the spherical-wave model~\eqref{channel model_LOS}. On the other hand, the subarray aperture for each subarray is far less than $\sqrt{\frac{\lambda D}{2}}$, since the antenna spacing is $0.5\lambda$, i.e., on the sub-millimeter level. Therefore, the channel elements within each subarray follow the similar expression with the planar-wave model~\eqref{channel_planar}. We denote $N_t$ and $N_r$ as the number of antennas at transmitter and receiver, respectively. 
By combining the spherical-wave propagation among subarrays and the planar-wave propagation within each subarray, the channel model of the THz WSMS architecture can be expressed as~\cite{11,8356240}
\begin{equation}	\textbf{H}=\sum\nolimits_{i=1}^{N_p}\alpha_i\textbf{G}_i\otimes(\textbf{a}_{ri}(\phi_{ri},\theta_{ri})\textbf{a}_{ti}(\phi_{ti},\theta_{ti})^H),
\label{channel_model_twolevel}
\end{equation}
where $\alpha_i$ denotes the channel gain of the $i^{\rm th}$ multipath and $N_p$ is the number of multipath. 
The $k\times k$-dimensional $\textbf{G}_i$ denotes the phase matrix among subarrays over the $i^{\rm th}$ multipath, satisfying $\textbf{G}_i[m,n]=e^{j\frac{2\pi}{\lambda}D^{mn}_{i}}$. $\otimes$ is the Kronecker product.
$\textbf{a}_{ri}(\phi_{ri},\theta_{ri})$ and $\textbf{a}_{ti}(\phi_{ti},\theta_{ti})$ are the array response vectors for the subarray of the $i^{\rm th}$ multipath at the receiver and transmitter, whose expressions are given in~\eqref{steering_vector}. $\phi_{ri} (\theta_{ri})$ and $\phi_{ti} (\theta_{ti})$ are the azimuth (elevation) angles of arrival and departure of the $i^{\rm th}$ multipath, respectively.
The phase model $\textbf{G}_i\otimes(\textbf{a}_{ri}(\phi_{ri},\theta_{ri})\textbf{a}_{ti}(\phi_{ti},\theta_{ti})^H)$ in channel model~\eqref{channel_model_twolevel} has been proved to be accurate in~\cite{8356240}. To make the whole channel model~\eqref{channel_model_twolevel} fit the THz band, we use the ray-tracing method presented in THz channel work~\cite{6998944} to calculate all parameters in~\eqref{channel_model_twolevel} to construct $\textbf{H}$.

\subsubsection{\textbf{Joint inter-path and intra-path multiplexing gain}}
We first present the following \textbf{\textit{Lemma 1}} to show the rank of channel $\textbf{H}$ of the THz WSMS architecture. 
Then, we analyze that the joint inter-path and intra-path multiplexing gain is $kN_p$ in the following \textit{\textbf{Remarks}}.

\textbf{\textit{Lemma 1:}} In the THz WSMS architecture, the rank of $\textbf{H}$ in~\eqref{channel_model_twolevel} is ${\rm min}\{{kN_p},{N_t},{N_r}\}$.

\textit{Proof:}
The rank of the $N_r\times N_t$-dimensional channel matrix $\textbf{H}$ is limited by the dimension, i.e., ${\rm min}\{N_t,N_r\}$. We first prove that when $kN_p\leq{\rm min}\{N_t,N_r\}$, the rank of $\textbf{H}$ equals to $kN_p$.
The channel matrix $\textbf{H}$ in \eqref{channel_model_twolevel} can be reformulated as 
\begin{subequations}
	\begin{align}
	\textbf{H}&\!=\!\sum\nolimits_{i=1}^{N_p}\alpha_i\textbf{G}_i\otimes\big(\textbf{a}_{ri}(\phi_{ri},\theta_{ri})\textbf{a}_{ti}(\phi_{ti},\theta_{ti})^H\big)
	\label{channel_proof_1}\\	
	&\!=\!\sum\nolimits_{i=1}^{N_p}\textbf{X}_i\otimes(\textbf{a}_{ri}\textbf{a}_{ti}^H)
	\label{channel_proof_2}\\
	&\!=\!\sum\nolimits_{i=1}^{N_p}(\textbf{I}_{k}\otimes\textbf{a}_{ri})(\textbf{X}_i\otimes\textbf{a}_{ti}^H)
	\label{channel_proof_3}\\
	&\!=\!\left[\textbf{I}_k\otimes \textbf{a}_{r1},...,\textbf{I}_k\otimes \textbf{a}_{rN_p}\right]\left[\begin{array}{c}
	\textbf{X}_1\otimes\textbf{a}_{t1}^H\\
	...\\
	\textbf{X}_{N_p}\otimes\textbf{a}_{tN_p}^H
	\end{array}\right]
	\label{channel_proof_4}
	\end{align}
\label{channel_proof}%
\end{subequations}
where $\textbf{X}_i=\alpha_{i}\textbf{G}_i$ in~\eqref{channel_proof_2}. The indices $(\phi_{ri},\theta_{ri})$ and  $(\phi_{ti},\theta_{ti})$ are omitted for simplicity in the following of this work. Step~\eqref{channel_proof_3} is the result from adopting the property of Kronecker product that $(\textbf{I}_k\textbf{X}_{i})\otimes(\textbf{a}_{ri}\textbf{a}_{ti}^H)=(\textbf{I}_{k}\otimes\textbf{a}_{ri})(\textbf{X}_i\otimes\textbf{a}_{ti}^H)$.
Step~\eqref{channel_proof_4} is the matrix form of~\eqref{channel_proof_3}. According to the structures of $\textbf{a}_{ri}$ and $\textbf{a}_{ti}$ in~\eqref{steering_vector}, one basic knowledge about $\textbf{a}_{ri}$ and $\textbf{a}_{ti}$ is that $\textbf{a}_{r1}$, ..., $\textbf{a}_{rN_p}$ are linearly independent and $\textbf{a}_{t1}$, ..., $\textbf{a}_{tN_p}$ are linearly independent, respectively. Therefore, it is straightforward to observe that the $N_r\times kN_p$-dimensional $\left[\textbf{I}_k\otimes \textbf{a}_{r1},...,\textbf{I}_k\otimes \textbf{a}_{rN_p}\right]$ is a column-full-rank matrix, whose rank is $kN_p$. Next, we prove that the $kN_p\times N_t$-dimensional $\big[
\textbf{X}_1\otimes\textbf{a}_{t1}^H;
...;
\textbf{X}_{N_p}\otimes\textbf{a}_{tN_p}^H
\big]$ is a row-full-rank matrix and the rank is $kN_p$. 

We use the case of $k=2$ as an example, while the general cases are similar and extensible.
We first assume that $\big[
\textbf{X}_1\otimes\textbf{a}_{t1}^H;
...;
\textbf{X}_{N_p}\otimes\textbf{a}_{tN_p}^H
\big]$ is NOT a row-full-rank matrix, which means that the row vectors of $\big[
\textbf{X}_1\otimes\textbf{a}_{t1}^H;
...;
\textbf{X}_{N_p}\otimes\textbf{a}_{tN_p}^H
\big]$ are not linearly independent. That is to say, $[x_1^{11},x_1^{12}]\otimes\textbf{a}_{t1}^H$, $[x_1^{21},x_1^{22}]\otimes\textbf{a}_{t1}^H$, ..., $[x_{N_p}^{11},x_{N_p}^{12}]\otimes\textbf{a}_{tN_p}^H$, $[x_{N_p}^{21},x_{N_p}^{22}]\otimes\textbf{a}_{tN_p}^H$ are not linearly independent, where $x_i^{mn}$ denotes the element at the $m^{\rm th}$ row and $n^{\rm th}$ column of $\textbf{X}_i$. Consequently, there exists a group of coefficients $\omega_i$ and $\gamma_i$ satisfying
\begin{equation}
\sum\nolimits_{i=1}^{N_p}\big(w_i[x_i^{11},x_i^{12}]\otimes\textbf{a}_{ti}^H+\gamma_i[x_i^{21},x_i^{22}]\otimes\textbf{a}_{ti}^H\big)=0,
\label{NON_LI_1}
\end{equation}
where $\omega_i\neq0$ and $\gamma_i\neq0$, $\exists i$.
\eqref{NON_LI_1} can be further expressed as
\begin{equation}
\sum\nolimits_{i=1}^{N_p}[w_ix_i^{11}+\gamma_ix_i^{21},w_ix_i^{12}+\gamma_ix_i^{22}]\otimes\textbf{a}_{ti}^H=0.
\label{NON_LI_2}
\end{equation}
Due to the linearly independent property of $\textbf{a}_{t1},...,\textbf{a}_{tN_p}$, \eqref{NON_LI_2} is equivalent to $[w_ix_i^{11}+\gamma_ix_i^{21},w_ix_i^{12}+\gamma_ix_i^{22}]=0$, $\forall i$. For the non-zero $w_i$ and $\gamma_i$, $[w_ix_i^{11}+\gamma_ix_i^{21},w_ix_i^{12}+\gamma_ix_i^{22}]=0$ is equal to
\begin{equation}
x_i^{11}x_i^{22}-x_i^{21}x_i^{12}=0.
\label{sufficient_condtion}
\end{equation}
Note that $x_i^{mn}=\textbf{X}_i[m,n]=\alpha_{i}\textbf{G}_i[m,n]$. $\textbf{G}_i$ is the phase matrix under the spherical-wave propagation, which has the similar form with $\textbf{G}$ in~\eqref{channel model_LOS} and is full-rank by avoiding the key-hole effect~\cite{Song2017,4155681}. Therefore, the determinant of $\alpha_{i}\textbf{G}_i$ is non-zero, i.e., $x_i^{11}x_i^{22}-x_i^{21}x_i^{12}\neq0$, which is contradictory with \eqref{sufficient_condtion}. Hence, the assumption that $\big[
\textbf{X}_1\otimes\textbf{a}_{t1}^H;
...;
\textbf{X}_{N_p}\otimes\textbf{a}_{tN_p}^H
\big]$ is not row-full-rank is invalid, i.e., $\big[
\textbf{X}_1\otimes\textbf{a}_{t1}^H;
...;
\textbf{X}_{N_p}\otimes\textbf{a}_{tN_p}^H
\big]$ is a row-full-rank matrix with the rank $2N_p$. Extending the case of $k=2$ to the general cases, $\big[
\textbf{X}_1\otimes\textbf{a}_{t1}^H;
...;
\textbf{X}_{N_p}\otimes\textbf{a}_{tN_p}^H
\big]$ is a row-full-rank matrix with the rank $kN_p$.

Until now, we have proved that in step~\eqref{channel_proof_4}, $\left[\textbf{I}_k\otimes \textbf{a}_{r1},...,\textbf{I}_k\otimes \textbf{a}_{rN_p}\right]$ is column-full-rank and $\big[
\textbf{X}_1\otimes\textbf{a}_{t1}^H;
...;
\textbf{X}_{N_p}\otimes\textbf{a}_{tN_p}^H
\big]$ is row-full-rank, respectively. Moreover, the rank of both these two matrices is $kN_p$. As a result, the rank of $\textbf{H}$ is $kN_p$.
When $kN_p$ exceeds ${\rm min}\{N_t,N_r\}$, owing to the dimension limitation of the $N_r\times N_t$-dimensional $\textbf{H}$, the rank of $\textbf{H}$ equals to ${\rm min}\{N_t,N_r\}$.
Hence, the rank of $\textbf{H}$ equals to ${\rm min}\{kN_p,N_t,N_r\}$, which completes the proof.
\hfill $\blacksquare$

\textit{\textbf{Remarks}}:
The multiplexing gain of the THz WSMS architecture equals to the rank of channel.
It is worth noting that when the number of subarrays $k$ is larger than ${\rm min}\{\frac{N_t}{k},\frac{N_r}{k}\}$, the spectral efficiency of THz WSMS architecture is significantly poor, which will be presented in Sec.~\ref{section_system_and_design}-B-1). Hence, we do not consider the case of $k>{\rm min}\{\frac{N_t}{k},\frac{N_r}{k}\}$ in this work and define the feasible set of $k$ as $\mathcal{K}$, which contains the integers from $1$ to ${\rm min}\{\frac{N_t}{k},\frac{N_r}{k}\}$. Therefore, the rank of $\textbf{H}$ equals to $kN_p$ such that \textbf{the joint inter-path and intra-path multiplexing gain of the WSMS architecture is} $\textit{\textbf{kN}}_\textit{\textbf{p}}$.
Specifically, the inter-path multiplexing gain coming from the multipath is $N_p$. The intra-path multiplexing gain provided by the spherical-wave propagation among the widely-spaced subarrays is $k$.
Compared to the existing THz hybrid beamforming architectures, the multiplexing gain of THz WSMS architecture is increased by $k$ times such that the spectral efficiency can be substantially improved.

\section{System Model and Problem Formulation of THz WSMS Architecture}
\label{section_system_and_design}
In this section, we investigate the system model and formulate a novel design problem for the THz WSMS hybrid beamforming architecture, which has not been studied yet. Different from the existing hybrid beamforming studies which aim to design the hybrid beamforming matrices to maximize the spectral efficiency, we additionally consider the optimization of the subarray spacing $d_s$ and the number of subarrays $k$.
Specifically, we first elaborate that $d_s$ and $k$ significantly influence the spectral efficiency of the THz WSMS hybrid beamforming architecture. Inspired by this, we formulate the design problem as jointly designing $d_s$, $k$, and hybrid beamforming matrices to maximize the spectral efficiency, which is then decomposed into two subproblems and solved efficiently.
\subsection{System Model}
As shown in Fig. \ref{architecture}, we assign $l_t$ and $l_r$ RF chains to control one subarray at the transmitter and receiver, respectively. The number of RF chains at the transmitter and receiver are $L_t=kl_t$ and $L_r=kl_r$. The number of data streams is $N_s$. To fully utilize the joint inter-path and intra-path multiplexing gain of the WSMS architecture, i.e., ${kN_p}$, we set $N_s=L_t=L_r=kN_p$. The system model of the THz WSMS hybrid beamforming is described as
\begin{equation}
\textbf{y} = \sqrt{\rho}\textbf{C}^{H}_{\rm D}\textbf{C}^{H}_{\rm A}\textbf{H}\textbf{P}_{\rm A}\textbf{P}_{\rm D}\textbf{s} +  \textbf{C}^{H}_{\rm D}\textbf{C}^{H}_{\rm A}\textbf{n},
\label{hybrid system model}
\end{equation}
where $\textbf{s}$ and $\textbf{y}$ are $N_s\times1$ transmitted and received signals, $\rho$ is the transmitted signal power. $\textbf{P}_{\rm A}\in\mathbb{C}^{N_t\times L_t}$ and $\textbf{P}_{\rm D}\in\mathbb{C}^{L_t\times N_s}$ are the analog and digital precoding matrices, respectively. $\textbf{C}_{\rm A}\in\mathbb{C}^{N_r\times L_r}$ and $\textbf{C}_{\rm D}\in\mathbb{C}^{L_r\times N_s}$ are the analog and digital combining matrices.
Moreover, $\textbf{n}\in\mathbb{C}^{N_r\times 1}$ represents the noise vector. 
Due to the hardware constraint that the subarrays are widely-spaced, the RF chain connected with one subarray can not connect to the other subarrays, which results in the block-diagonal structure of $\textbf{P}_{\rm A}$ as
\begin{equation}
\setlength{\arraycolsep}{1pt} 
\textbf{P}_{\rm A}=\left[\begin{array}{ccccccccc}
\textbf{p}_{11}&...&\textbf{p}_{1l_t}&\bm{0}&...&...&...&...&\bm{0}\\
\bm{0}&...&\bm{0}&\textbf{p}_{21}&...&\textbf{p}_{2l_t}&\bm{0}&...&\bm{0}\\
...&...&...&...&...&...&...&...&...\\
\bm{0}&...&...&...&...&\bm{0}&\textbf{p}_{k1}&...&\textbf{p}_{kl_t}\end{array}\right],
\label{structure of PA DAoSA}
\end{equation}
where $\textbf{p}_{il}$ is an $\frac{N_{t}}{k} \times 1$ vector and $\bm{0}$ is an $\frac{N_{t}}{k} \times 1$ zero vector. Since $\textbf{p}_{il}$ is implemented by phase shifters, the elements of $\textbf{p}_{il}$ follow the constant modulus (CM) constraint, i.e., $\lvert\textbf{p}_{il}[m]\rvert^2=1$, $\forall m$. The structure and constraints of $\textbf{C}_{\rm A}$ are the same as $\textbf{P}_{\rm A}$ in~\eqref{structure of PA DAoSA} and $\textbf{c}_{il}$ is an $\frac{N_{r}}{k} \times 1$ vector. Additionally, the transmitter's power constraint is given by $\lVert\textbf{P}_{\rm A}\textbf{P}_{\rm D}\rVert^{2}_{F}=N_s$.
Therefore, the achievable spectral efficiency of the THz WSMS hybrid beamforming architecture is
\begin{equation}
SE={\rm{log}_{2}}\Big(\big\lvert \textbf{I}_{N_s}+\frac{\rho}{N_{s}}\textbf{R}_{n}^{\!-1}\textbf{C}^{H}_{\rm D}\textbf{C}^{H}_{\rm A}\textbf{H}\textbf{P}_{\rm{A}}\textbf{P}_{\rm{D}}\textbf{P}^{H}_{\rm{D}}
\textbf{P}^{H}_{\rm{A}}\textbf{H}^{H}\textbf{C}_{\rm A}\textbf{C}_{\rm D}\big\rvert\Big),
\label{R}
\end{equation}
where $\textbf{R}_{n}=\sigma_{n}^{2}\textbf{C}_{\rm D}^{H}\textbf{C}_{\rm A}^{H}\textbf{C}_{\rm A}\textbf{C}_{\rm D}$ is the noise covariance matrix after combining, and $\sigma_{n}^2$ denotes the noise power. Note that a deep CNN-powered channel estimation method is proposed in~\cite{9322174} to acquire the channel state information (CSI) for $\textbf{H}$ in~\eqref{channel_model_twolevel}, where the channel estimation normalized-mean-square-error is below -15 dB, i.e., the near-optimal CSI can be obtained. Hence, in this work, we assume that full-CSI has been obtained.

\subsection{Problem Formulation}
In the existing hybrid beamforming studies~\cite{1,7397861,DAoSA_JSAC_2020,7445130,7389996,9411813}, the design problem is usually formulated as designing $\textbf{P}_{\rm A}$, $\textbf{P}_{\rm D}$, $\textbf{C}_{\rm A}$, and $\textbf{C}_{\rm D}$ to maximize the spectral efficiency. While in the THz WSMS architecture, besides $\textbf{P}_{\rm A}$, $\textbf{P}_{\rm D}$, $\textbf{C}_{\rm A}$, and $\textbf{C}_{\rm D}$, the subarray spacing $d_s$ and the number of subarrays $k$ also have significant impact on spectral efficiency, which need to be carefully designed. Therefore, we first elaborate the influence of $d_s$ and $k$ on spectral efficiency and reveal that it is necessary to optimize $d_s$ and $k$. Then, we formulate the design problem of WSMS architecture as jointly designing $d_s$, $k$, $\textbf{P}_{\rm A}$, $\textbf{P}_{\rm D}$, $\textbf{C}_{\rm A}$, and $\textbf{C}_{\rm D}$ to maximize the spectral efficiency. 

\subsubsection{\textbf{Impact of $d_s$ and $k$ on spectral efficiency}}
To mitigate the influence of $\textbf{P}_{\rm A}$, $\textbf{P}_{\rm D}$, $\textbf{C}_{\rm A}$, and $\textbf{C}_{\rm D}$ when focusing on the impact of $d_s$ and $k$, we consider that the optimal precoding and combining are conducted such that the spectral efficiency equals to the capacity and is not related to $\textbf{P}_{\rm A}$, $\textbf{P}_{\rm D}$, $\textbf{C}_{\rm A}$, and $\textbf{C}_{\rm D}$. It is worth noting that, we indeed propose the optimal closed-form solution for precoding and combining matrices in Sec.~\ref{section_beamforming} such that this consideration is reasonable.

Note that from \textbf{\textit{Lemma 1}} and \textbf{\textit{Remarks}} in Sec.~\ref{section_channel_CS_WS}-C, the rank of channel $\textbf{H}$ is $kN_p$. 
Considering that the optimal precoding and combining are conducted, the spectral efficiency of WSMS architecture equals to the capacity, which is given by
\begin{equation}
\begin{aligned}
SE_{\rm optimal}=\sum\nolimits_{i=1}^{kN_p}{\rm{log}_{2}}\Big( 1+\frac{\rho_i}{\sigma_{n}^2}r_i^2(\textbf{H})\Big),
\label{capacity_1}
\end{aligned}
\end{equation}
where $r_i(\textbf{H})$ is the $i^{\rm th}$ largest singular value of $\textbf{H}$. Let $z^+\triangleq{\rm max}\{z,0\}$, $\rho_i=\Big(\Gamma-\frac{\sigma_{n}^2}{r_i^2(\textbf{H})}\Big)^+$ describes the transmitted power allocated to the $i^{\rm th}$ sub-channel, where $\Gamma$ is chosen to satisfy the transmitted power constraint $\sum_{i}\rho_i=\rho$. One property about $r_i(\textbf{H})$ is that $\sum_{i}r_i^2(\textbf{H})=\lVert\textbf{H}\rVert_{F}^2$. The energy of channel $\lVert\textbf{H}\rVert_{F}^2$ depends on the strength of each multipath and the number of antennas. When $N_t$ and $N_r$ are fixed, for THz WSMS architecture with varying $k$ and $d_s$, the array aperture is usually on the order of $\sqrt{\frac{\lambda D}{2}}$ and much smaller than the communication distance $D$, as we analyzed in Sec.~\ref{section_channel_CS_WS}-C. As a result, the strength of each multipath is almost unchanged such that $\lVert\textbf{H}\rVert_{F}^2$ is a constant for THz WSMS architecture with different $d_s$ and $k$.

\begin{figure}
	\centering
	\captionsetup{font={footnotesize}}
	\subfigure[Spectral efficiency versus $d_s$.]{	\includegraphics[scale=0.55]{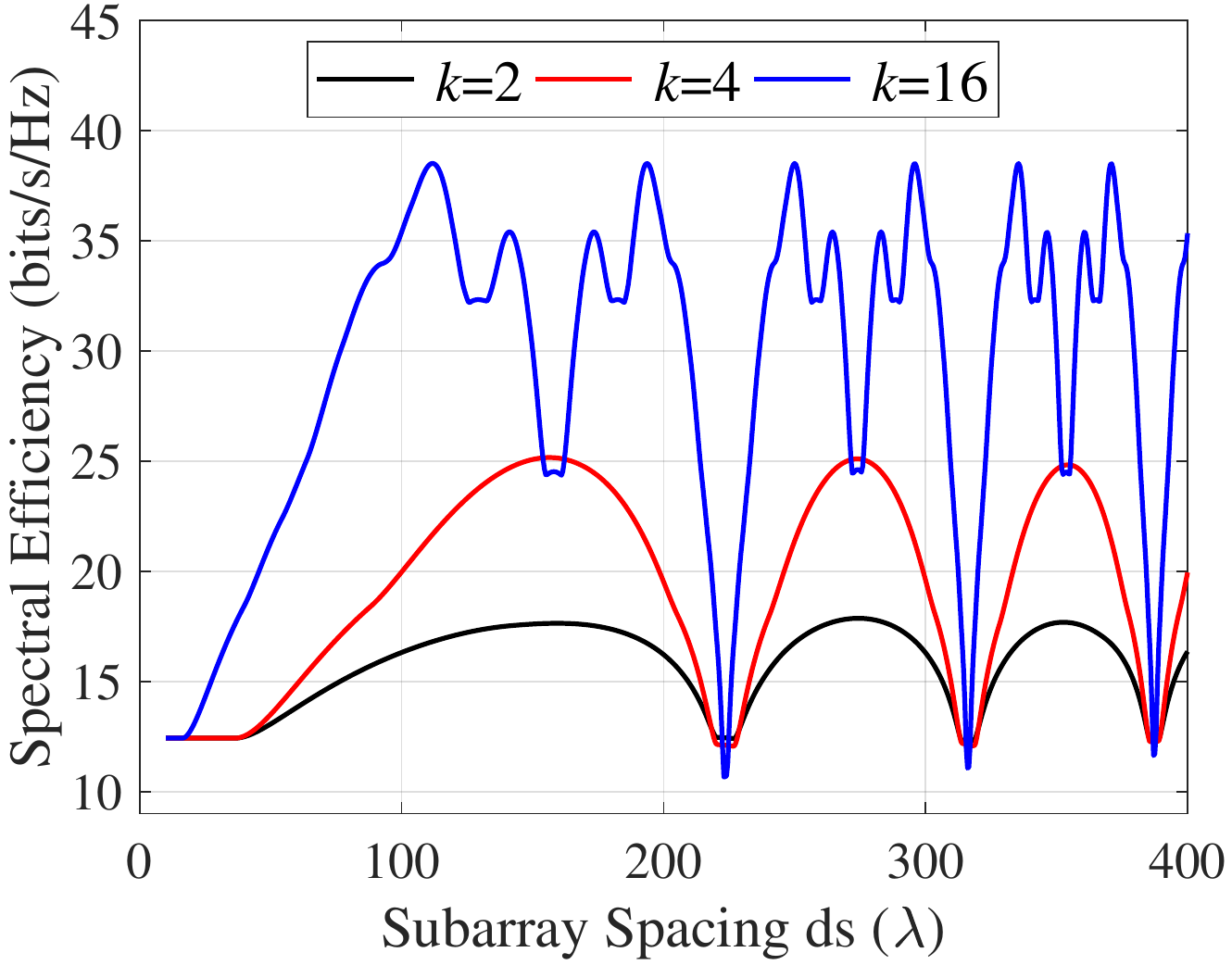}}
	\subfigure[Spectral efficiency versus $k$.]{\includegraphics[scale=0.55]{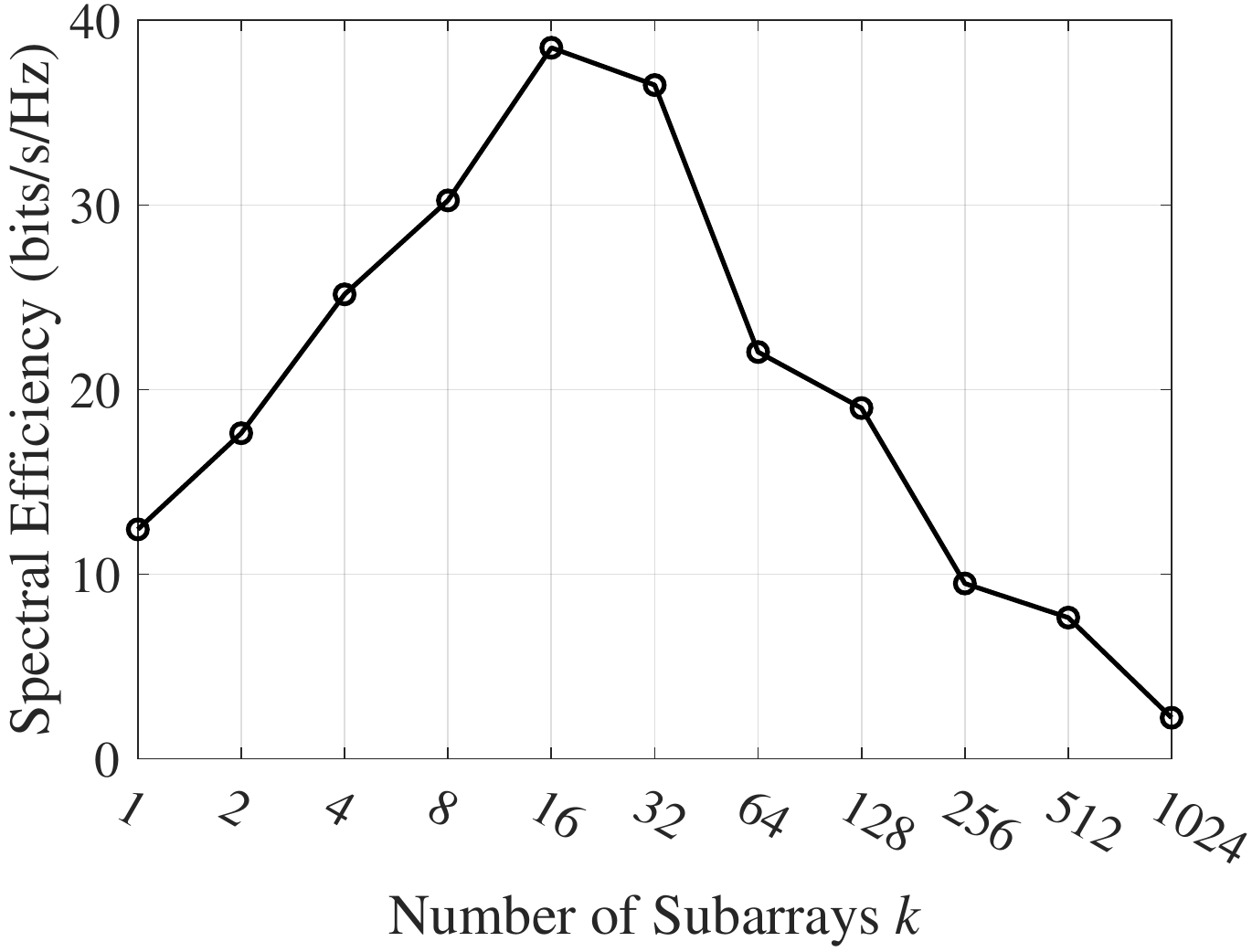}}
	\caption{The spectral efficiency of the THz WSMS architecture versus $d_s$ and $k$. $\rho=10$ dBm. $N_t=N_r=1024$.}
	\label{Figure_C_versus_ds_k}
	\vspace{-7.5mm}
\end{figure}

{\textbf{Spectral efficiency versus $d_s$}:} As analyzed in~\eqref{channel_model_twolevel}, with different subarray spacing $d_s$, $D_i^{mn}$ changes such that the phase term $\textbf{G}_i$ in channel $\textbf{H}$ changes, which further influences the spectral efficiency of the THz WSMS architecture. 
We evaluate the spectral efficiency of THz WSMS architecture versus $d_s$ numerically in an exemplary application of 0.3 THz wireless backhaul with 5 GHz bandwidth. The transmitter and receiver are two base stations with heights of $h_t=h_r=30$m, $D=50$m. Due to the long communication distance and high reflection and scattering loss, one LoS path and one ground-reflection path are considered~\cite{AKYILDIZ201416,doi10106315014037}. As shown in Fig.~\ref{Figure_C_versus_ds_k}(a), for each $k$, the spectral efficiency of WSMS architecture varies rapidly with $d_s$. Hence, the optimization of $d_s$ is critical to the spectral efficiency.

{\textbf{Spectral efficiency versus $k$}:} As shown in~\eqref{capacity_1}, spectral efficiency of the THz WSMS architecture equals to the summation of spectral efficiencies of $kN_p$ equivalent sub-channels, whose equivalent channel gain is $r_i(\textbf{H})$.

1) With larger $k$, the number of sub-channels increases, which enhances the spectral efficiency~\eqref{capacity_1}.

2) With larger $k$ and more sub-channels, the equivalent channel gain $r_i(\textbf{H})$ and corresponding SNR of each sub-channel decrease, since the summation of the square of equivalent channel gain $\sum_{i}r_i^2(\textbf{H})=\lVert\textbf{H}\rVert_{F}^2$ is a constant, as analyzed above. This fact reduces the spectral efficiency~\eqref{capacity_1}.

We evaluate the spectral efficiency versus $k$ numerically in Fig.~\ref{Figure_C_versus_ds_k}(b). For each point of $k$, $d_s$ is set as the optimal value as in Fig.~\ref{Figure_C_versus_ds_k}(a). In Fig.~\ref{Figure_C_versus_ds_k}(b), the spectral efficiency first increases and then decreases with $k$. The reason is that when $k$ is small, the increase of the number of sub-channels dominates the trend of spectral efficiency. While when $k$ is large, the reduction of equivalent channel gain of each sub-channel becomes more important. Consequently, the optimization of $k$ is critical to the spectral efficiency. We also show the spectral efficiency of THz WSMS architecture with $k>{\rm min}\{\frac{N_t}{k},\frac{N_r}{k}\}$, i.e., $k=1024$, in~Fig.~\ref{Figure_C_versus_ds_k}(b). The spectral efficiency is significantly poor due to the low equivalent channel gain of each sub-channel. Therefore, we do not consider the case of $k>{\rm min}\{\frac{N_t}{k},\frac{N_r}{k}\}$ in this work.

\subsubsection{\textbf{Design problem}} To sum up, the spectral efficiency of WSMS architecture highly depends on $d_s$ and $k$, which motivates the optimization of $d_s$ and $k$. Moreover, hybrid beamforming matrices $\textbf{P}_{\rm A}$, $\textbf{P}_{\rm D}$, $\textbf{C}_{\rm A}$, and $\textbf{C}_{\rm D}$ are critical to the spectral efficiency of the WSMS architecture. Consequently, we formulate our design problem as jointly designing $d_s$, $k$, $\textbf{P}_{\rm A}$, $\textbf{P}_{\rm D}$, $\textbf{C}_{\rm A}$, and $\textbf{C}_{\rm D}$ to maximize the spectral efficiency~\eqref{R} of the THz WSMS hybrid beamforming architecture as
{\setlength\abovedisplayskip{-0.5cm}
	\begin{subequations}
	\begin{align}
	{\mathop{\rm \ max}\limits_{d_s,k,\textbf{P}_{\rm A},\textbf{P}_{\rm D},\textbf{C}_{\rm A},\textbf{C}_{\rm D}}}&{\rm{log}_{2}}\Big(\big\lvert \textbf{I}_{N_s}+\frac{\rho}{N_{s}}\textbf{R}_{n}^{\!-1}\textbf{C}^{H}_{\rm D}\textbf{C}^{H}_{\rm A}\textbf{H}\textbf{P}_{\rm{A}}\textbf{P}_{\rm{D}}\textbf{P}^{H}_{\rm{D}}
	\textbf{P}^{H}_{\rm{A}}\textbf{H}^{H}\textbf{C}_{\rm A}\textbf{C}_{\rm D}\big\rvert\Big)
	\label{fff_obj}\\ 
	&\mathrm{s.t.}\  \lvert\textbf{p}_{il}[m]\rvert^2=1, \forall i,l,m
	\label{fff_cons_1}\\
	&\quad\ \ \lvert\textbf{c}_{il}[m]\rvert^2=1, \forall i,l,m
	\label{fff_cons_2}\\
	&\quad\ \
	\lVert \textbf{P}_{\rm A}\textbf{P}_{\rm D}\rVert^{2}_{F}=N_{s}
	\label{fff_cons_3}\\
	&\quad\ \ d_{s}^{\rm min} \leq d_{s}\leq d_{s}^{\rm max}, k\in\mathcal{K},
	\label{fff_cons_4}
	\end{align}
	\label{fff}%
\end{subequations}} 
where~\eqref{fff_cons_1} and~\eqref{fff_cons_2} come from the CM constraint in~\eqref{structure of PA DAoSA}. \eqref{fff_cons_3} is the transmitter's power constraint.
$d_{s}^{\rm max}$ denotes the maximal subarray spacing, which is decided by the maximal array aperture by considering the practical communication system. The minimal subarray spacing $d_{s}^{\rm min}$ is defined to avoid overlap of adjacent subarrays. $\mathcal{K}$ is the feasible set of $k$, which contains the integers from $1$ to ${\rm min}\{\frac{N_t}{k},\frac{N_r}{k}\}$, as analyzed in Sec.~\ref{section_channel_CS_WS}-C-2).

Directly solving this problem is intractable since the variables $d_s$, $k$, $\textbf{P}_{\rm A}$, $\textbf{P}_{\rm D}$, $\textbf{C}_{\rm A}$, and $\textbf{C}_{\rm D}$ are coupled together.
Hence, to make the problem tractable to solve, we decompose the design problem into two subproblems. The first subproblem P1 is a hybrid beamforming problem, i.e., we calculate $\textbf{P}_{\rm A}$, $\textbf{P}_{\rm D}$, $\textbf{C}_{\rm A}$, and $\textbf{C}_{\rm D}$ to maximize~\eqref{fff_obj}, by assuming that $d_s$ and $k$ are given and known. The second subproblem P2 is an array configuration problem, i.e., we determine $d_s$ and $k$ to maximize~\eqref{fff_obj}. As a result, by solving P1 and P2, the design problem~\eqref{fff} can be solved. Next, we propose an optimal closed-form solution for the hybrid beamforming problem P1, in Sec.~\ref{section_beamforming}. In Sec.~\ref{section_design_subarray}, we propose a DLR algorithm to solve the array configuration problem P2. 

\section{Optimal Closed-form Solution for Hybrid Beamforming Problem P1}
\label{section_beamforming}
In this section, we propose the optimal closed-form solution to the hybrid beamforming problem P1, through exploiting the structure of channel of THz WSMS architecture. The objective of P1 is designing $\textbf{P}_{\rm A}$, $\textbf{P}_{\rm D}$, $\textbf{C}_{\rm A}$, and $\textbf{C}_{\rm D}$ to maximize the spectral efficiency~\eqref{fff_obj}, given $d_s$ and $k$. The constraints of P1 are~\eqref{fff_cons_1},~\eqref{fff_cons_2}, and~\eqref{fff_cons_3}.

It is well-known that to solve P1 optimally, we need to design $\textbf{P}_{\rm A}$, $\textbf{P}_{\rm D}$, $\textbf{C}_{\rm A}$, and $\textbf{C}_{\rm D}$ to satisfy $\textbf{P}_{\rm opt}=\textbf{P}_{\rm A}\textbf{P}_{\rm D}$ and $\textbf{C}_{\rm opt}=\textbf{C}_{\rm A}\textbf{C}_{\rm D}$, respectively. By defining the SVD of $\textbf{H}$ as $\textbf{H}=\textbf{U}\bm{\Sigma}\textbf{V}^H$, $\textbf{C}_{\rm opt}=\textbf{U}_{N_s}$ is the first $N_s$ columns of $\textbf{U}$.
$\textbf{P}_{\rm opt}$ equals to $\textbf{V}_{N_s}\bm{\Gamma}$, where $\textbf{V}_{N_s}$ refers to the first $N_s$ columns of $\textbf{V}$ and $\bm{\Gamma}$ is the water-filling power allocation matrix.

However, using most of the existing hybrid beamforming algorithms~\cite{7397861,DAoSA_JSAC_2020,7389996,7445130}, it is intractable to find $\textbf{P}_{\rm A}$, $\textbf{P}_{\rm D}$, $\textbf{C}_{\rm A}$, and $\textbf{C}_{\rm D}$ to satisfy $\textbf{P}_{\rm opt}=\textbf{P}_{\rm A}\textbf{P}_{\rm D}$ and $\textbf{C}_{\rm opt}=\textbf{C}_{\rm A}\textbf{C}_{\rm D}$ due to the following reasons. First is the non-convex constraints~\eqref{fff_cons_1} and~\eqref{fff_cons_2}. Second is the block-diagonal structure of $\textbf{P}_{\rm A}$ and $\textbf{C}_{\rm A}$ in~\eqref{structure of PA DAoSA}, which results in that some elements of $\textbf{P}_{\rm A}$ and $\textbf{C}_{\rm A}$ are 0 and can not be designed. We propose an optimal closed-form solution of $\textbf{P}_{\rm A}$, $\textbf{P}_{\rm D}$, $\textbf{C}_{\rm A}$, and $\textbf{C}_{\rm D}$ as follows. We first propose the optimal $\textbf{P}_{\rm A}$ and $\textbf{P}_{\rm D}$ and then propose the optimal $\textbf{C}_{\rm A}$ and $\textbf{C}_{\rm D}$. To start, we decompose the channel matrix $\textbf{H}$ in~\eqref{channel_model_twolevel} as follows.
\begin{subequations}
	\begin{align}
	\textbf{H}&=\sum\nolimits_{i=1}^{N_p}\alpha_i\textbf{G}_i\otimes\big(\textbf{a}_{ri}\textbf{a}_{ti}^H\big)\label{eq_H_transform_first_step}\\
	&=\left[\begin{array}{ccc}{\sum\nolimits_{i=1}^{N_p}\alpha_i\textbf{G}_i[1,1]\textbf{a}_{ri}\textbf{a}_{ti}^H}&{\cdots}&{\sum\nolimits_{i=1}^{N_p}\alpha_i\textbf{G}_i[1,k]\textbf{a}_{ri}\textbf{a}_{ti}^H} \\{\vdots}&{\ddots}&{\vdots}\\{\sum\nolimits_{i=1}^{N_p}\alpha_i\textbf{G}_i[k,1]\textbf{a}_{ri}\textbf{a}_{ti}^H}&{\cdots}&{\sum\nolimits_{i=1}^{N_p}\alpha_i\textbf{G}_i[k,k]\textbf{a}_{ri}\textbf{a}_{ti}^H}\end{array}\right]\label{eq_H_transform_1}\\
	&=\left[\begin{array}{ccc}{\textbf{A}_{r}\bm{\Lambda}_{11}\textbf{A}_{t}^H}&{\cdots}&{\textbf{A}_{r}\bm{\Lambda}_{1k}\textbf{A}_{t}^H} \\{\vdots}&{\ddots}&{\vdots}\\{\textbf{A}_{r}\bm{\Lambda}_{k1}\textbf{A}_{t}^H}&{\cdots}&{\textbf{A}_{r}\bm{\Lambda}_{kk}\textbf{A}_{t}^H}\end{array}\right]\label{eq_H_transform_2}\\
	&=\left[\begin{array}{ccc}{\textbf{A}_{r}\bm{\Lambda}_{11}}&{\cdots}&{\textbf{A}_{r}\bm{\Lambda}_{1k}} \\{\vdots}&{\ddots}&{\vdots}\\{\textbf{A}_{r}\bm{\Lambda}_{k1}}&{\cdots}&{\textbf{A}_{r}\bm{\Lambda}_{kk}}\end{array}\right]\cdot\big(\textbf{I}_{k}\otimes \textbf{A}_{t}^H\big)\label{eq_H_transform_3}\\
	&=\textbf{B}\big(\textbf{I}_{k}\otimes \textbf{A}_{t}^H\big),
	\label{eq_H_transform}
	\end{align}
\end{subequations}
where $(\phi_{ri},\theta_{ri})$ and $(\phi_{ti},\theta_{ti})$ of $\textbf{a}_{ri}$ and $\textbf{a}_{ti}^H$ are omitted for simplicity. In~\eqref{eq_H_transform_2}, $\textbf{A}_{r}=[\textbf{a}_{r1},...,\textbf{a}_{rN_p}]$, $\textbf{A}_{t}=[\textbf{a}_{t1},...,\textbf{a}_{tN_p}]$, and $\bm{\Lambda}_{mn}={\rm diag}(\alpha_1\textbf{G}_i[m,n],...,\alpha_{N_p}\textbf{G}_{N_p}[m,n])$. \eqref{eq_H_transform_3} comes from the property of the product of block matrix.

As we analyzed before, through the SVD, $\textbf{H}$ equals to $\textbf{U}\bm{\Sigma}\textbf{V}^{H}$. According to~\eqref{eq_H_transform}, we have $\textbf{B}\big(\textbf{I}_{k}\otimes \textbf{A}_{t}^H\big)=\textbf{U}\bm{\Sigma}\textbf{V}^{H}$. Using the unitary property of $\textbf{U}$ that $\textbf{U}\textbf{U}^H=\textbf{U}^H\textbf{U}=\textbf{I}_{N_r}$, $\textbf{B}\big(\textbf{I}_{k}\otimes \textbf{A}_{t}^H\big)=\textbf{U}\bm{\Sigma}\textbf{V}^{H}$ can be derived as $\bm{\Sigma}^{-1}\textbf{U}^H\textbf{B}\big(\textbf{I}_{k}\otimes \textbf{A}_{t}^H\big)=\textbf{V}^{H}$, which further implies
\begin{subequations}
	\begin{align}
	&\textbf{V}=\big(\textbf{I}_{k}\otimes \textbf{A}_{t}\big)\textbf{B}^H\textbf{U}(\bm{\Sigma}^{-1})^H=\big(\textbf{I}_{k}\otimes \textbf{A}_{t}\big)\textbf{T},\\
	&\textbf{V}_{N_s}=\big(\textbf{I}_{k}\otimes \textbf{A}_{t}\big)\textbf{T}_{N_s},	
	\end{align}
	\label{expression_T}%
\end{subequations}
where $\textbf{T}=\textbf{B}^H\textbf{U}(\bm{\Sigma}^{-1})^H$ and $\textbf{T}_{N_s}$ is the first $N_s$ columns of $\textbf{T}$. In addition, $\textbf{V}_{N_s}$ is the first $N_s$ columns of $\textbf{V}$, as we analyzed before.
As a result, the optimal precoding matrix $\textbf{P}_{\rm opt}=\textbf{V}_{N_s}\bm{\Gamma}$ can be represented as
\begin{equation}
\textbf{P}_{\rm opt}=\textbf{V}_{N_s}\bm{\Gamma}=\underbrace{\big(\textbf{I}_{k}\otimes \textbf{A}_{t}\big)}_{\textbf{P}_{\rm A}}\cdot \underbrace{\textbf{T}_{N_s}\bm{\Gamma}}_{\textbf{P}_{\rm D}}.
\label{solution_precoding}
\end{equation}
As shown in~\eqref{structure of PA DAoSA}, the constraints of the $N_t\times L_t$-dimensional $\textbf{P}_{\rm A}$ are the block-diagonal constraint and the CM constraint. Interestingly, we observe that $\textbf{I}_{k}\otimes \textbf{A}_{t}$ is an $N_t\times kN_p$-dimensional matrix, which follows the block-diagonal and the CM constraints in~\eqref{structure of PA DAoSA}. It is worth noting that in WSMS architecture, we have $L_t=L_r=N_s=kN_p$ to fully utilize the joint inter-path and intra-path multiplexing gain. Therefore, $\textbf{I}_{k}\otimes \textbf{A}_{t}$ has the same dimension with $\textbf{P}_{\rm A}$ and satisfies the constraints of $\textbf{P}_{\rm A}$. Consequently, $\textbf{I}_{k}\otimes \textbf{A}_{t}$ is indeed the optimal solution of $\textbf{P}_{\rm A}$. The dimension of $\textbf{T}_{N_s}\bm{\Gamma}$ is $kN_p\times N_s$, which equals to the dimension of the digital precoding matrix $\textbf{P}_{\rm D}$, i.e., $L_t\times N_s$. Since there is no constraint on $\textbf{P}_{\rm D}$, we can use $\textbf{T}_{N_s}\bm{\Gamma}$ as the optimal solution of $\textbf{P}_{\rm D}$. As a result, the proposed solutions of $\textbf{P}_{\rm A}$ and $\textbf{P}_{\rm D}$ in~\eqref{solution_precoding} satisfy the optimal condition $\textbf{P}_{\rm opt}=\textbf{P}_{\rm A}\textbf{P}_{\rm D}$.
Using the similar procedures from~\eqref{eq_H_transform_first_step} to~\eqref{eq_H_transform}, $\textbf{H}$ can also be expressed as 
\begin{equation}
	\textbf{H}=\big(\textbf{I}_{k}\otimes \textbf{A}_{r}\big)\cdot\left[\!\!\!\!\begin{array}{ccc}{\bm{\Lambda}_{11}\textbf{A}_{t}^H}&{\cdots}&{\bm{\Lambda}_{1k}\textbf{A}_{t}^H} \\{\vdots}&{\ddots}&{\vdots}\\{\bm{\Lambda}_{k1}\textbf{A}_{t}^H}&{\cdots}&{\bm{\Lambda}_{kk}\textbf{A}_{t}^H}\end{array}\!\!\!\right]=\big(\textbf{I}_{k}\otimes \textbf{A}_{r}\big)\textbf{D}.
\end{equation}
Therefore, we have $\big(\textbf{I}_{k}\otimes \textbf{A}_{r}\big)\textbf{D}=\textbf{U}\bm{\Sigma}\textbf{V}^{H}$, which further implies that $\textbf{U}=\big(\textbf{I}_{k}\otimes \textbf{A}_{r}\big)\textbf{D}\textbf{V}\bm{\Sigma}^{-1}=\big(\textbf{I}_{k}\otimes \textbf{A}_{r}\big)\textbf{R}$, where $\textbf{R}\!=\textbf{D}\textbf{V}\bm{\Sigma}^{-1}$, according to the unitary property of $\textbf{V}$ that $\textbf{V}\textbf{V}^H=\textbf{V}^H\textbf{V}=\textbf{I}_{N_t}$. Consequently, the optimal combining matrix $\text{C}_{\rm opt}=\textbf{U}_{N_s}$ can be expressed as
\begin{equation}
\text{C}_{\rm opt}=\textbf{U}_{N_s}=\underbrace{\big(\textbf{I}_{k}\otimes \textbf{A}_{r}\big)}_{\textbf{C}_{\rm A}}\cdot \underbrace{\textbf{R}_{N_s}}_{\textbf{C}_{\rm D}},
\label{solution_combining}
\end{equation}
where $\textbf{R}_{N_s}$ is the first $N_s$ columns of $\textbf{R}$. Similar to $\textbf{P}_{\rm A}$ and $\textbf{P}_{\rm D}$, it is straightforward to observe that $\textbf{I}_{k}\otimes \textbf{A}_{r}$ and $\textbf{R}_{N_s}$ are the optimal solutions of $\textbf{C}_{\rm A}$ and $\textbf{C}_{\rm D}$. The pseudo codes to implement the optimal closed-form solution of P1 are described in \textbf{Algorithm 1}.
\begin{table}
	\centering
	\begin{tabular}{p{280pt}}
		\hline \textbf{Algorithm 1: Optimal closed-form hybrid beamforming solution for P1} \\
		\hline \textbf{Input:} Parameters of $\textbf{H}$ in~\eqref{channel_model_twolevel}, $k$, $d_s$\\
		\quad1:\ Construct $\textbf{H}$, calculate $\bm{\Gamma}$, $\bm{\Lambda}_{11}$, $\bm{\Lambda}_{12}$, ..., $\bm{\Lambda}_{kk}$\\
		\quad2:\ According to~\eqref{expression_T} and~\eqref{solution_combining}, construct $\textbf{T}_{N_s}$ and $\textbf{R}_{N_s}$ \\			
		\quad3:\ $\textbf{P}_{\rm A}=\textbf{I}_k\otimes \textbf{A}_t$ and $\textbf{P}_{\rm D}=\textbf{T}_{N_s}\bm{\Gamma}$ \\
		\quad4:\ $\textbf{C}_{\rm A}=\textbf{I}_k\otimes \textbf{A}_r$ and $\textbf{C}_{\rm D}=\textbf{R}_{N_s}$ \\
		\quad5:\ Normalize $\textbf{P}_{\rm D}$ as $\textbf{P}_{\rm D}=\frac{\sqrt{N_s}}{\ \ \left\lVert\textbf{P}_{\rm A}{\textbf{P}_{\rm D}}\right\rVert_{F}}{\textbf{P}_{\rm D}}$ to satisfy~\eqref{fff_cons_3}  \\				
		\textbf{Output:} $\textbf{P}_{\rm A}$, $\textbf{P}_{\rm D}$, $\textbf{C}_{\rm A}$, and $\textbf{C}_{\rm D}$\\
		\hline
	\end{tabular}
	\vspace{-6.5mm}
\end{table} 

Since the proposed optimal closed-form hybrid beamforming solution to P1 only requires basic operations of matrices and does not involve iteration, the computational complexity is as low as $\mathcal{O}((N_t+N_r)N_s^2)$, which is linearly related with the number of antennas $N_t$ and $N_r$. On the contrary, the computational complexities of most existing hybrid beamforming algorithms are proportional to the square or even higher order of $N_t$ and $N_r$~\cite{7397861,DAoSA_JSAC_2020,7445130,7389996,8356240}.

\section{DLR Algorithm for Array Configuration Problem P2}
\label{section_design_subarray}
In this section, we propose a low-complexity DLR algorithm to solve the array configuration problem P2, i.e., designing $d_s$ and $k$ to maximize the spectral efficiency of the THz WSMS architecture. The key idea of the DLR algorithm is utilizing the dominant LoS property of the THz band to relax the intractable P2 into a tractable problem.

In Sec.~\ref{section_beamforming}, we have proposed the optimal $\textbf{P}_{\rm A}$, $\textbf{P}_{\rm D}$, $\textbf{C}_{\rm A}$, and $\textbf{C}_{\rm D}$ to maximize the spectral efficiency. By substituting the optimal $\textbf{P}_{\rm A}$, $\textbf{P}_{\rm D}$, $\textbf{C}_{\rm A}$, and $\textbf{C}_{\rm D}$ in~\eqref{fff_obj}, the spectral efficiency equals to the capacity in~\eqref{capacity_1}. As a result, problem P2 is equivalent to designing $d_s$ and $k$ to maximize the capacity of THz WSMS architecture as follows. 
\begin{subequations}
	\begin{align}
	{\rm P2:}\quad {\mathop{\rm \ max}\limits_{d_{s}, k}}& \sum\nolimits_{i=1}^{kN_p}{\rm{log}_{2}}\Big( 1+\frac{\rho_i}{\sigma_{n}^2}r_i^2(\textbf{H})\Big)
	\label{C_obj}
	\\ 
	\mathrm{s.t.}\ &{\rho}_i=\Big({\Gamma}-\frac{\sigma_{n}^2}{r_i^2(\textbf{H})}\Big)^+, \sum\nolimits_{i=1}^{k}{\rho}_i=\rho \label{C_cons_1}\\
	&d_{s}^{\rm min} \leq d_{s}\leq d_{s}^{\rm max}, k\in\mathcal{K},
	\label{C_cons_2}
	\end{align}
	\label{P1}%
\end{subequations}
where~\eqref{C_cons_1} comes from~\eqref{capacity_1}. \eqref{C_cons_2} denotes the constraints of $d_s$ and $k$, as analyzed in~\eqref{fff_cons_4}.

A straightforward and optimal method for solving P2 is using exhaustive search to select $d_{s}$ and $k$. However, this is impractical since the number of candidates of $d_{s}$ is prohibitively large. 
Instead, to design $k$ and $d_s$ efficiently, we propose a DLR algorithm as follows.
The joint design of $k$ and $d_s$ is difficult, since these two parameters are coupled in $\textbf{H}$. Therefore, we first shed light on the design of the subarray spacing $d_{s}$ to maximize~\eqref{C_obj}, with a fixed value of $k$. After that, we determine the number of subarrays $k$ to maximize~\eqref{C_obj}.
\subsection{Design of the Subarray Spacing $d_s$ in DLR Algorithm}
\begin{figure}
	\centering
	\captionsetup{font={footnotesize}}
	\includegraphics[scale=0.55]{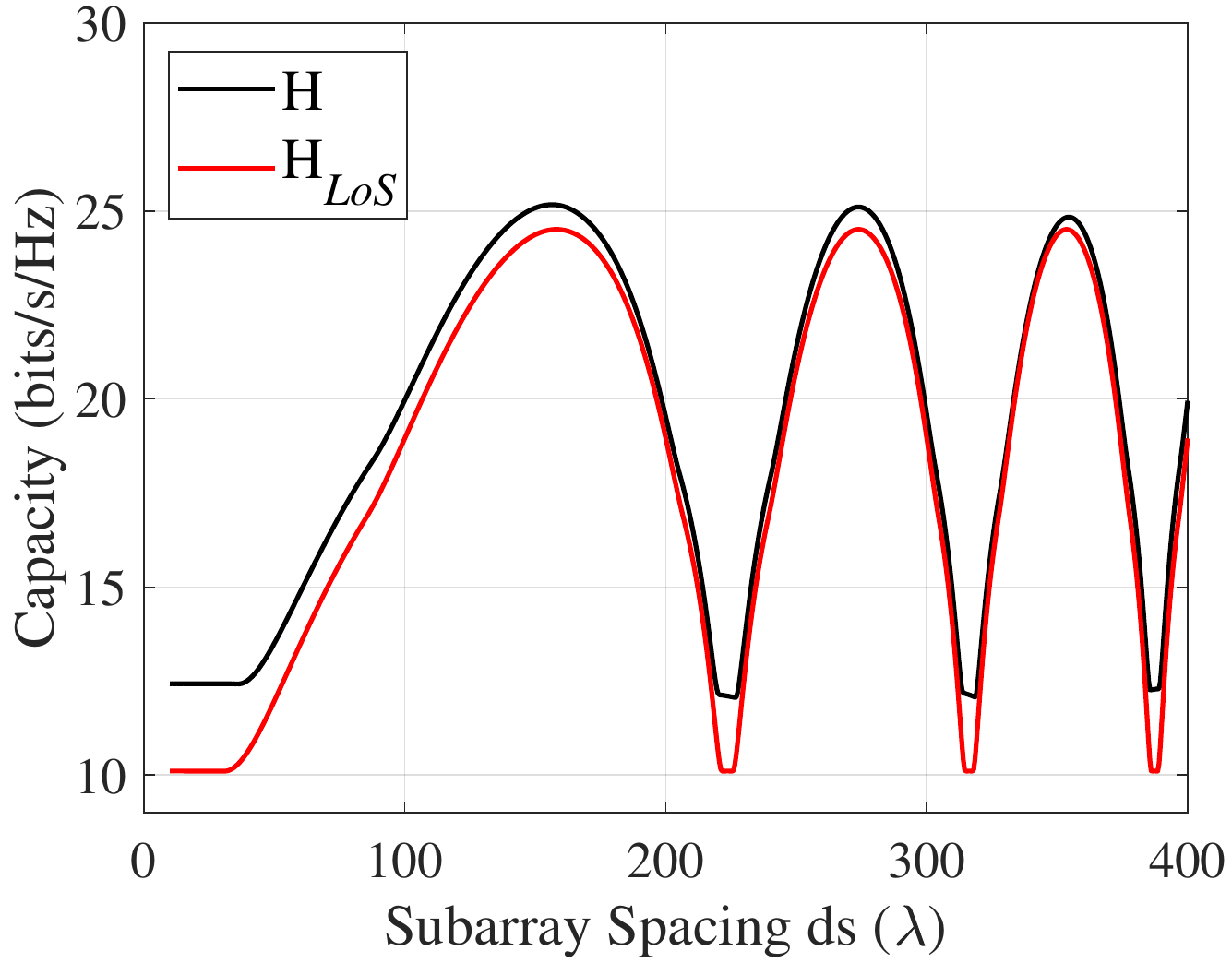}
	\caption{Capacities of $\textbf{H}$ and $\textbf{H}_{LoS}$ versus $d_{s}$. $N_t=N_r=1024$, $k=4$, $D=50$m, $h_t=h_r=30$m, $\rho=10$ dBm.}
	\label{Figure_CapacityVersusSpacing}
	\vspace{-6.5mm}
\end{figure}

To take a closer look, the channel matrix $\textbf{H}$ in \eqref{channel_model_twolevel} can be divided into two parts as 
\begin{align}
\textbf{H}=\textbf{H}_{LoS}+\textbf{H}_{NLoS}=\alpha_1\textbf{G}_1\otimes(\textbf{a}_{r1}\textbf{a}_{t1}^H)+\sum\nolimits_{i=2}^{N_p}\alpha_i\textbf{G}_i\otimes(\textbf{a}_{ri}\textbf{a}_{ti}^H),
\label{channel_model_twolevel_LoS}
\end{align}
where $\textbf{H}_{LoS}$ and $\textbf{H}_{NLoS}$ refer to the LoS and NLoS parts of $\textbf{H}$, respectively.
Compared to the microwave and mmWave frequencies, one peculiarity of the THz channel is the dominant LoS property, i.e., the received power through the LoS path is more than $10$ times stronger than the other NLoS paths~\cite{8387210,7786122}. 
Inspired by this property, we propose to maximize the capacity of $\textbf{H}_{LoS}$, instead of the capacity of $\textbf{H}$. To show the feasibility of this substitution, we compare the capacities of $\textbf{H}_{LoS}$ and $\textbf{H}$ versus $d_{s}$ in Fig.~\ref{Figure_CapacityVersusSpacing}. 
The setup of the channel follows the analysis in Sec.~\ref{section_system_and_design}-B-1). 
Since the capacity of $\textbf{H}_{LoS}$ contributes to about 97\% capacity of the entire channel $\textbf{H}$, the $d_s$ which maximizes the capacity of $\textbf{H}_{LoS}$ will also provide a near-optimal solution to maximize the capacity of $\textbf{H}$, with deviations less than 2\%.
Therefore, it is reasonable to design $d_s$ to maximize the capacity of $\textbf{H}_{LoS}$, instead of directly maximizing~\eqref{C_obj}.

The rank of $\textbf{H}_{LoS}$ equals to $kN_p$ with $N_p=1$, i.e., equals to $k$. Thus, the capacity of $\textbf{H}_{LoS}$ is
\begin{equation}
C_{LoS}=\sum\nolimits_{i=1}^{k}{\rm{log}_{2}}\Big( 1+\frac{\widehat{\rho}_i}{\sigma_{n}^2}r_i^2(\textbf{H}_{LoS})\Big),
\label{C_LoS}
\end{equation}
where $\widehat{\rho}_i=\Big(\widehat{\Gamma}-\frac{\sigma_{n}^2}{r_i^2(\textbf{H}_{LoS})}\Big)^+$ is the    water-filling transmitted power allocated to the $i^{\rm th}$ sub-channel, $\widehat{\Gamma}$ is chosen to satisfy the transmitted power constraint $\sum_i\widehat{\rho}_i=\rho$, and $r_{i}(\textbf{H}_{LoS})$ denotes the $i^{\rm th}$ largest singular value of $\textbf{H}_{LoS}$. 

Designing $d_s$ to maximize~\eqref{C_LoS} is still ambitious due to the complicated water-filling power allocation $\widehat{\rho}_i$. To make the optimization tractable, we substitute the water-filling allocation with equal-power allocation $\widehat{\rho}_i=\rho/k$, which yields a lower bound of $C_{LoS}$. As a result, the design of $d_s$ is relaxed as maximizing the lower bound of $C_{LoS}$ as
\begin{equation}
{\mathop{\rm \ max}\limits_{d_{s}}} \sum\nolimits_{i=1}^{k}{\rm{log}_{2}}\Big( 1+\frac{\rho}{k\sigma_{n}^2}r_i^2(\textbf{H}_{LoS})\Big).
\label{C_LOS_2_label_1}	
\end{equation}
Following the Jensen's inequality, \eqref{C_LOS_2_label_1} can be rearranged as
\begin{equation}
    \sum\nolimits_{i=1}^{k}{\rm{log}_{2}}\Big(1+\frac{\rho}{k\sigma_{n}^2}r_i^2(\textbf{H}_{LoS})\Big)
	\leq k\cdot{\rm{log}_{2}}\Big(1+\frac{1}{k}\cdot\frac{\rho}{k\sigma_{n}^2}\sum\nolimits_{i=1}^{k}r_i^2(\textbf{H}_{LoS})\Big),
	\label{capacity_Jensen}
\end{equation}
where the right hand of~\eqref{capacity_Jensen} is a constant since $\sum_{i=1}^{k}r^2_i(\textbf{H}_{LoS})=\lVert\textbf{H}_{LoS}\rVert_{F}^2$ is a constant when we designing $d_s$, as analyzed in Sec.~\ref{section_system_and_design}-B-1).
Consequently, the right hand of~\eqref{capacity_Jensen} is the upper bound of~\eqref{C_LOS_2_label_1} such that designing $d_s$ to maximize~\eqref{C_LOS_2_label_1} can be transformed into designing $d_s$ to make the equality in~\eqref{capacity_Jensen} hold. According to Jensen's inequality, the equality holds when all $r_{i}(\textbf{H}_{LoS})$ are identical. 
Recall that $\textbf{H}_{LoS}=\alpha_1\textbf{G}_1\otimes(\textbf{a}_{r1}\textbf{a}_{t1}^H)$.
According to the property of Kronecker product, $r_{1}(\textbf{H}_{LoS})=r_{2}(\textbf{H}_{LoS})=\ldots=r_{k}(\textbf{H}_{LoS})$ is equivalent to $r_{1}(\textbf{G}_{1})=r_{2}(\textbf{G}_{1})=\ldots=r_{k}(\textbf{G}_{1})$, since $\textbf{a}_{r1}\textbf{a}_{t1}^H$ only has one singular value.
As analyzed in \eqref{channel_model_twolevel}, the structure of $\textbf{G}_{1}\textbf{G}_{1}^H$ is
\begin{equation}
\setlength{\arraycolsep}{1pt} 
\textbf{G}_{1}\textbf{G}_{1}^H\!\!=\!\!\left[\begin{array}{ccc}{k}&{\cdots}&{\sum\nolimits_{i=1}^{k}} e^{j\frac{2\pi}{\lambda}\left(D^{1i}_1\!-\!D^{k i}_1\right)} \\{\vdots}&{\ddots}&{\vdots}\\{\sum\nolimits_{i=1}^{k}e^{j \frac{2\pi}{\lambda}\left(D^{ki}_1\!-\!D^{1i}_1\right)}}&{\cdots}&{k}\end{array}\right]\!.
\label{structure_G1G1H}
\end{equation}
Following the property of singular value and the structure of $\textbf{G}_{1}\textbf{G}_{1}^H$, $r_{1}(\textbf{G}_{1})=r_{2}(\textbf{G}_{1})=\ldots=r_{k}(\textbf{G}_{1})$ is further equivalent to $\textbf{G}_{1}\textbf{G}_{1}^H=k\textbf{I}_{k}$. Therefore, the design of $d_{s}$ can be transformed into making the difference between $\textbf{G}_1\textbf{G}_1^H$ and $k\textbf{I}_{k}$ as small as possible, i.e., to minimize $\lVert\textbf{G}_{1}\textbf{G}_{1}^H-k\textbf{I}_{k}\rVert_{F}^2$.
According to the structure of $\textbf{G}_1\textbf{G}_1^H$ in \eqref{structure_G1G1H}, $\lVert\textbf{G}_{1}\textbf{G}_{1}^H-k\textbf{I}_{k}\rVert_{F}^2$ can be rewritten as
\begin{equation}
\lVert\textbf{G}_{1}\textbf{G}_{1}^H-k\textbf{I}_{k}\rVert_{F}^2=-k^3+\sum\limits_{a=1}^{k}\sum\limits_{b=1}^{k}
\underbrace{\Big\lvert{\sum\limits_{i=1}^{k}} e^{j\frac{2\pi}{\lambda}\left(D^{ai}_1\!-\!D^{b i}_1\right)}\Big\rvert^2}_{(\star)}.
\label{eq:f_d_s}
\end{equation}
As we showed in Sec.~\ref{section_channel_CS_WS}-C-1), $\frac{2\pi}{\lambda}D^{mn}_1$ can be represented as
\begin{equation}
\label{eq:Dmn1}
\frac{2\pi}{\lambda}D^{mn}_{1}=\frac{2\pi}{\lambda}\sqrt{D^2+(l^{mn})^2},
\end{equation}
where $(l^{mn})^2=((x_m-x_n)d_{s})^2+((z_m-z_n)d_{s})^2$. The square root form brings difficulties to the calculation. Therefore, we propose to use the Taylor expansion to simplify the expression of $\frac{2\pi}{\lambda}D^{mn}_1$ as
\begin{align}
\frac{2\pi}{\lambda}D^{mn}_1=\frac{2\pi D}{\lambda}\sqrt{1+\Big(\frac{l^{mn}}{D}\Big)^2}=\frac{2\pi D}{\lambda}\Big(1+\frac{1}{2}\Big(\frac{l^{mn}}{D}\Big)^2-\frac{1}{8}\Big(\frac{l^{mn}}{D}\Big)^4+\ldots\Big).
\end{align}
Noting that $l^{mn}$ ranges from $0$ to the array aperture $S$, which is on the order of $\sqrt{\frac{\lambda D}{2}}$. As a result, the maximal value of $\frac{2\pi D}{\lambda}(\frac{-1}{8}(\frac{l^{mn}}{D})^4)$ and the higher order terms of the Taylor expansion are on the order of $\frac{\lambda}{D}$, which can be omitted since the wavelength $\lambda$ of THz wave is much smaller than the communication distance $D$. Hence, $D^{mn}_1$ in \eqref{eq:Dmn1} can be rearranged as 
\begin{equation}
\label{eq:Dmn2}
D^{mn}_1\approx   D+\frac{d_s^2}{2D}((x_m-x_n)^2+(z_m-z_n)^2).
\end{equation}
By substituting \eqref{eq:Dmn2} in the expression $(\star)$ in \eqref{eq:f_d_s}, we can further derive 
\begin{subequations}
	\begin{align}
	\Big\lvert{\sum\limits_{i=1}^{k}} e^{j\frac{2\pi}{\lambda}\left(D^{ai}_1\!-\!D^{b i}_1\right)}\Big\rvert^2
	&=\!\Big(\!\sum_{i=1}^{k}\!{\rm cos}\big(\frac{2\pi}{\lambda}(D^{ai}_1\!-\!D^{bi}_1)\big)\!\Big)^2\!\!+\!\Big(\!\sum_{i=1}^{k}\!{\rm sin}\big(\frac{2\pi}{\lambda}(D^{ai}_1\!-\!D^{bi}_1)\big)\!\Big)^2
	\label{Taylor_expasion_cons1}\\
	&=k\!+\!\!\sum_{i=1}^{k}\sum_{l=i+1}^{k}\!\!2{\rm cos}\Big(\frac{2\pi}{\lambda}\big((D^{ai}_1\!-\!D^{bi}_1)\!-\!(D^{al}_1\!-\!D^{bl}_1)\big)\Big)
	\label{Taylor_expasion_cons2}\\
	&\approx k\!+\!\!\sum_{i=1}^{k}\sum_{l=i+1}^{k}\!\!2{\rm cos}\Big(\frac{2\pi d_s^2}{\lambda D}\psi_{a,b,i,l}\Big),
	\label{Taylor_expasion_cons3}
	\end{align}
	\label{Taylor_expasion}%
\end{subequations}
where \eqref{Taylor_expasion_cons3} is the result of substituting \eqref{eq:Dmn2} in \eqref{Taylor_expasion_cons2} and $\psi_{a,b,i,l}=(x_a-x_b)(x_l-x_i)+(z_a-z_b)(z_l-z_i)$.	By replacing the expression $(\star)$ in \eqref{eq:f_d_s} with \eqref{Taylor_expasion_cons3}, the problem in~\eqref{C_LOS_2_label_1} can be reformulated as 
\begin{equation}
\begin{aligned}
{\mathop{\rm \ min \ }\limits_{d_{s}}} f(d_s&)=\sum\limits_{a=1}^{k}\sum\limits_{b=1}^{k}\sum_{i=1}^{k}\sum_{l=i+1}^{k}2{\rm cos}\Big(\frac{2\pi d_s^2}{\lambda D}\psi_{a,b,i,l}\Big)\\ 
\mathrm{s.t.}\ &d_{s}^{\rm min} \leq d_{s}\leq d_{s}^{\rm max}.
\end{aligned}
\label{C_LOS_4}
\end{equation}
To now, the intractable problem P2 is transferred into a tractable problem~\eqref{C_LOS_4}. Since the objective function of~\eqref{C_LOS_4} has an explicit expression about the variable $d_s$, we invoke the gradient descend method to calculate $d_s$ to solve~\eqref{C_LOS_4}. 


%
%
%

\subsection{Design of the Number of Subarrays $k$ in DLR Algorithm}
Different from the subarray spacing that has a prohibitively large number of candidate values, the possible value of subarrays $k$ is very limited. 
As we analyzed before, the feasible set $\mathcal{K}$ of $k$ contains the integers from $1$ to ${\rm min}\{\frac{N_t}{k},\frac{N_r}{k}\}]$.
For instance, when $N_t=N_r=1024$, $k$ can only be selected from $\mathcal{K}=\{1,2,4,8,16,32,64,128,256,512\}$,
Consequently, the exhaustive search on $k$ has a reasonably low complexity.

\begin{table}
	\centering
	\begin{tabular}{p{300pt}}
		\hline \textbf{Algorithm 2: DLR algorithm for problem P2} \\
		\hline \textbf{Input:} $D$, parameters of $\textbf{H}$ in~\eqref{channel_model_twolevel} except for $d_s$ and $k$\\
		\quad1:\ Using gradient descend method to solve~\eqref{C_LOS_4} to obtain $d_s$, for each $k\in\mathcal{K}$ \\
		\quad2:\ For each pair of $d_s$ and $k$, construct $\textbf{H}$ and calculate~\eqref{C_obj} \\			
		\quad3:\ Select the $d_s$ and $k$ which have largest~\eqref{C_obj} \\
		\textbf{Output:} $d_s$ and $k$\\
		\hline	
	\end{tabular}
	\vspace{-5.5mm}
\end{table} 
\begin{table}
	\centering
	\begin{tabular}{p{300pt}}
		\hline \textbf{Algorithm 3: Total algorithm to solve the overall design problem~\eqref{fff}} \\
		\hline \textbf{Input:} $D$, parameters of $\textbf{H}$ in~\eqref{channel_model_twolevel} except for $d_s$ and $k$\\
		\quad1:\ Run Algorithm 2 to obtain $d_s$ and $k$ \\
		\quad2:\ Run Algorithm 1 to obtain $\textbf{P}_{\rm A}$, $\textbf{P}_{\rm D}$, $\textbf{C}_{\rm A}$, and $\textbf{C}_{\rm D}$ \\			
		\textbf{Output:} $d_s$, $k$, $\textbf{P}_{\rm A}$, $\textbf{P}_{\rm D}$, $\textbf{C}_{\rm A}$, and $\textbf{C}_{\rm D}$\\
		\hline	
	\end{tabular}
	\vspace{-6.5mm}
\end{table}

The pseudo codes to implement the DLR algorithm are described in \textbf{Algorithm 2}. Moreover, the flowchart about using the optimal closed-form hybrid beamforming solution and the DLR algorithm to solve the overall design problem~\eqref{fff} is presented in \textbf{Algorithm 3}.

\textbf{\textit{Discussions:}}
By invoking the DLR algorithm, we design $d_s$ and $k$ to maximize the spectral efficiency of the THz WSMS hybrid beamforming architecture.
For practical implementation of the array, the subarray spacing $d_s$ and the number of subarrays $k$ are fixed and can not be adjusted after the time of manufacture. This implies that we need to determine $d_s$ and $k$ according to DLR algorithm before manufacturing. As shown in~\textbf{Algorithm 2}, the communication distance $D$ is one input of the DLR algorithm. Hence, we consider the following two types of applications. On one hand, for scenarios with fixed transmitter and receiver, e.g., THz wireless backhaul, $D$ is a fixed value such that we can directly put this $D$ as the input of DLR algorithm and design $d_s$ and $k$. On the other hand, while for mobile scenarios, $D$ is a random variable with the probability $p(D=D_q)=p_q$. When leveraging the DLR algorithm, the objective function~\eqref{C_LOS_4} needs to be changed to $\sum_{q}p_qf(d_s,D_q)$ to consider all the possible $D$. Similarly, \eqref{C_obj} needs changes as~\eqref{C_LOS_4}. To evaluate the performance of the THz WSMS hybrid beamforming architecture with the proposed algorithms in this work, we focus on the THz wireless backhaul scenario.

\section{Performance Evaluation}
\label{section_simulation}
In this section, we extensively evaluate the spectral efficiency as well as the energy efficiency of the THz WSMS hybrid beamforming architecture with the proposed algorithms. The counterparts in existing studies are presented in TABLE~\ref{Table_counterpart}. Particularly, compared to~\cite{8356240}, the major novelties of the WSMS architecture in this work focus on two points. First, we propose a DLR algorithm to optimize $d_s$ and $k$ to maximize the spectral efficiency. While in~\cite{8356240}, only a special case that $k=2$ and $d_s\approx \sqrt{\lambda D/2}$ is considered and the optimization of $k$ and $d_s$ is missed.
Second, regarding the optimization of $\textbf{P}_{\rm A}$, $\textbf{P}_{\rm D}$, $\textbf{C}_{\rm A}$, and $\textbf{C}_{\rm D}$, we propose a closed-form solution, whose performance is optimal and the computational complexity is substantially lower than the hybrid beamforming algorithm in~\cite{8356240}.

\begin{table}
	\centering
	\captionsetup{font={footnotesize}}
	\caption{Proposed scheme in this work versus the existing studies.}
	\begin{tabular}{|m{51pt}<{\centering}|m{167pt}<{\centering}|m{117pt}<{\centering}|m{71pt}<{\centering}|}
		\hline
		Architecture&Design problem&Applied algorithms&Multiplexing gain\\
		\hline
		THz WSMS in this work&\textbf{Jointly design} $d_s$, $k$, $\textbf{P}_{\rm A}$, $\textbf{P}_{\rm D}$, $\textbf{C}_{\rm A}$, $\textbf{C}_{\rm D}$ to maximize the spectral efficiency (SE)&\textbf{Proposed optimal closed-form solution} and \textbf{DLR algorithm}&Joint inter-path and intra-path: $ \textit{\textbf{kN}}_\textit{\textbf{p}}$\\
		\hline
		THz WSMS in~\cite{8356240}&Design  $\textbf{P}_{\rm A}$, $\textbf{P}_{\rm D}$, $\textbf{C}_{\rm A}$, $\textbf{C}_{\rm D}$ to maximize SE, for the special case $k=2$ and $d_s\approx\sqrt{\lambda D/2}$&Algorithm in~\cite{8356240}&Joint inter-path and intra-path: $2N_p$\\
		\hline
		FC in~\cite{7397861}&Design $\textbf{P}_{\rm A}$, $\textbf{P}_{\rm D}$, $\textbf{C}_{\rm A}$, $\textbf{C}_{\rm D}$ to maximize SE&PE-AltMin algorithm&Inter-path: $N_p$\\
		\hline
		AoSA in~\cite{7445130}&Design $\textbf{P}_{\rm A}$, $\textbf{P}_{\rm D}$, $\textbf{C}_{\rm A}$, $\textbf{C}_{\rm D}$ to maximize SE&SIC algorithm&Inter-path: $N_p$\\
		\hline
		LoS MIMO in~\cite{8466787}&Design fully-digital precoding and combining matrices to maximize SE& Unconstrained digital beamforming algorithm&Intra-path: ${\rm min}\{N_t,N_r\}$\\
		\hline
	\end{tabular}
	\label{Table_counterpart}
	\vspace{-7.5mm}
\end{table}

\textbf{Simulation setup}: We consider a THz wireless backhaul scenario~\cite{AKYILDIZ201416,doi10106315014037}. The height of the transmitter and receiver is $h_t=h_r=30$m. The communication distance $D$ ranges from $60$m, $70$m, $80$m, $90$m, to $100$m.
We consider a LoS path and a ground-reflection path, whose path gains can be obtained by the ray-tracing method~\cite{8387210,6998944}. 
The central frequency is $f=0.3$ THz. The bandwidth $B$ is 5 GHz and the noise power is -76.2~dBm. For THz WSMS architecture, we set $N_s=L_t=L_r=kN_p$ to fully utilize the joint inter-path and intra-path multiplexing gain. For FC and AoSA architectures, we set $N_s=L_t=L_r=N_p$ to fully harvest the inter-path multiplexing gain. The LoS MIMO architecture is a fully-digital architecture such that $L_t=N_t$, $L_r=N_r$, and $N_s={\rm min}\{N_t, N_r\}$.
\subsection{Spectral Efficiency Comparison}
\begin{figure}
	\centering
	\begin{tabular}{cc}
		\begin{minipage}[t]{0.45\linewidth}
			\includegraphics[width = 1\linewidth]{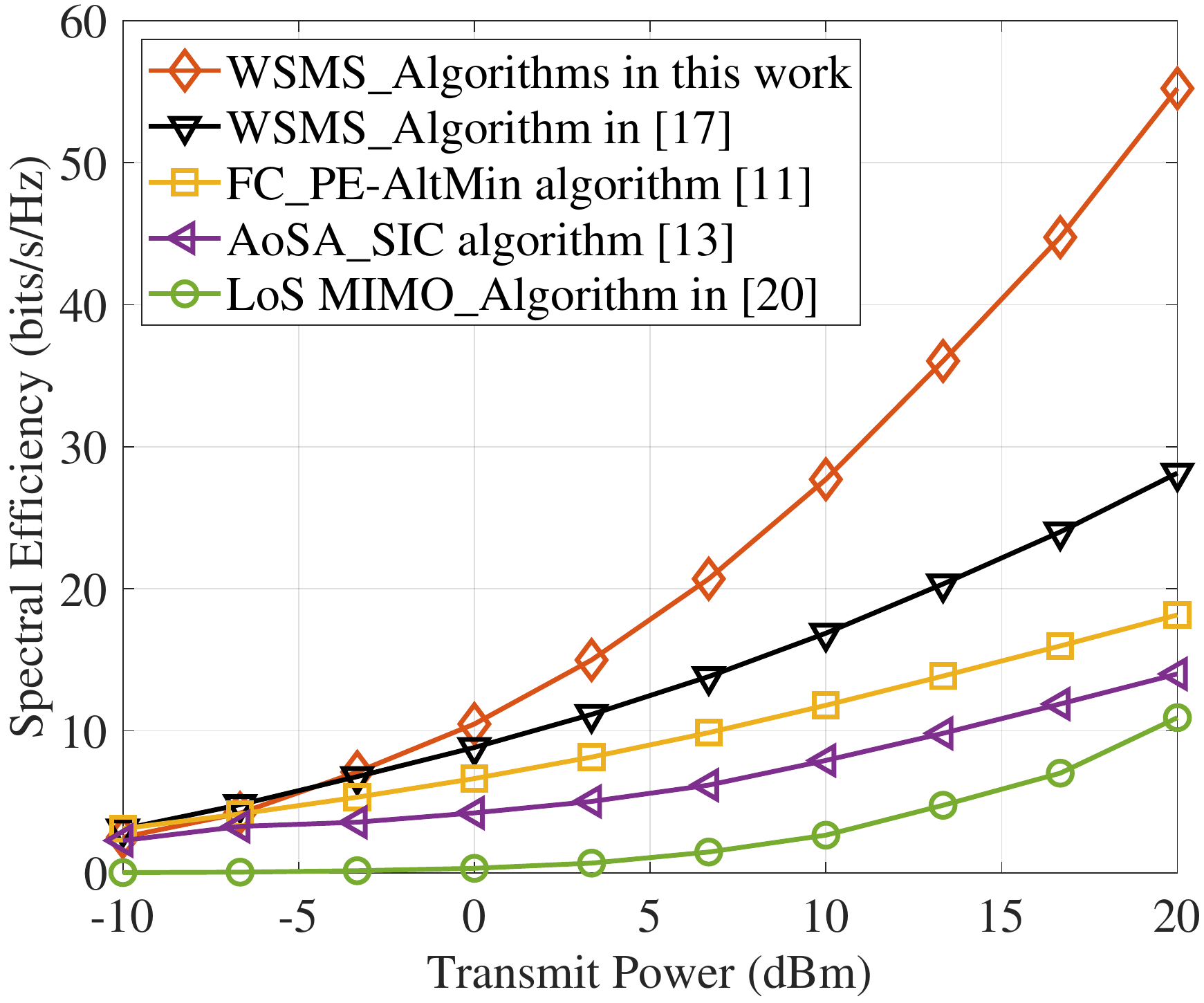}
			\captionsetup{font={footnotesize}}
			\caption{Spectral efficiency versus transmit power $\rho$, $N_t=N_r=1024$, $D=60$m.}
			\label{fig_SE_vs_rho}
		\end{minipage}
		\begin{minipage}[t]{0.45\linewidth}
			\includegraphics[width = 1\linewidth]{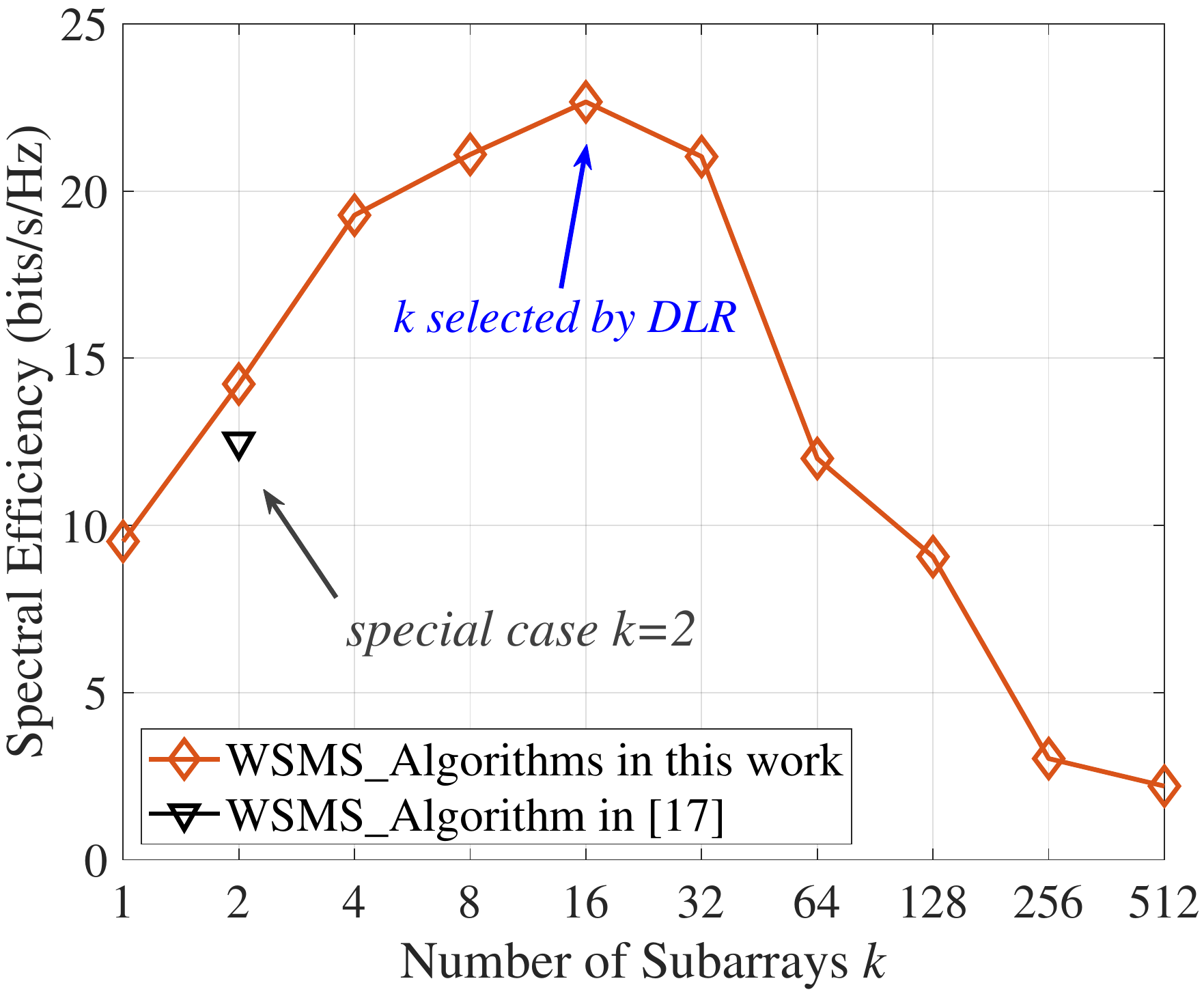}
			\captionsetup{font={footnotesize}}
			\caption{Spectral efficiency versus $k$, $N_t=N_r=1024$, $D=80$m, $\rho=$ 10 dBm.}
			\label{fig_SE_vs_NtNr}
		\end{minipage}
	\end{tabular}
	\vspace{-9.5mm}
\end{figure}
Fig.~\ref{fig_SE_vs_rho} shows the spectral efficiency of the THz WSMS architecture versus transmit power $\rho$. Due to the joint inter-path and intra-path multiplexing, the total multiplexing gain of THz WSMS architecture is $kN_p$, which is $k$ times of the multiplexing gain of the FC and AoSA architectures with only inter-path multiplexing. As a result, the spectral efficiency of THz WSMS architecture is much higher than the FC and AoSA architectures, e.g., 189\% and 358\% higher when $\rho=20$ dBm. Particularly, using our proposed algorithms in this paper, the spectral efficiency of WSMS architecture is much higher than WSMS architecture with algorithm in~\cite{8356240}, e.g., 96\% higher when $\rho=20$ dBm. The reasons are illustrated as follows. 
On one hand, in this work, we propose a DLR algorithm to optimize $k$ and $d_s$ to maximize the spectral efficiency of the WSMS architecture. However, the WSMS architecture in~\cite{8356240} only considers a special case that $k=2$ and $d_s\approx \sqrt{\lambda D/2}$ without optimization of $k$ and $d_s$, which restricts the spectral efficiency. On the other hand, in this work, we obtain the optimal $\textbf{P}_{\rm A}$, $\textbf{P}_{\rm D}$, $\textbf{C}_{\rm A}$, and $\textbf{C}_{\rm D}$ to maximize the spectral efficiency, through the proposed optimal closed-form solution. While $\textbf{P}_{\rm A}$, $\textbf{P}_{\rm D}$, $\textbf{C}_{\rm A}$, and $\textbf{C}_{\rm D}$ designed by the algorithm in~\cite{8356240} are non-optimal, which reduce the spectral efficiency.
Last, the LoS MIMO architecture achieves a very low spectral efficiency due to the following reasons. As analyzed in Sec.~\ref{section_channel_CS_WS}-B, the intra-path multiplexing gain in LoS MIMO architecture equals to the number of antennas, i.e., 1024.
It is well known that, through SVD, the channel matrix $\textbf{H}_{\rm intra}$ of LoS MIMO architecture can be decomposed into 1024 sub-channels, where the equivalent channel gain of the $i^{\rm th}$ sub-channel is the the $i^{\rm th}$ singular value. The summation of the square of singular value is the energy of the whole channel $\lVert\textbf{H}_{\rm intra}\rVert_{F}^2$. Dividing $\lVert\textbf{H}_{\rm intra}\rVert_{F}^2$ into 1024 parts, the equivalent channel gain of each sub-channel is very poor. Therefore, although with abundant 1024 intra-path multiplexing gain, the poor equivalent channel gain of each sub-channel leads to a low spectral efficiency for LoS MIMO architecture. 

Fig.~\ref{fig_SE_vs_NtNr} further demonstrates that the spectral efficiency of THz WSMS architecture with the proposed algorithms in this work is much higher than the THz WSMS architecture with the algorithm in~\cite{8356240}. As shown in Fig.~\ref{fig_SE_vs_NtNr}, $k$ has a significant impact on spectral efficiency. Using DLR algorithm, we select $k=16$ to achieve the highest spectral efficiency. While for WSMS architecture in~\cite{8356240}, only a special case $k=2$ is considered and the spectral efficiency is much lower than the WSMS architecture with $k=16$ in this work.
Moreover, even with the same $k=2$, the spectral efficiency of WSMS architecture in this work is higher than the WSMS architecture in~\cite{8356240}. The reason comes from two aspects. First, in this work, $d_s$ is also optimized by DLR algorithm to maximize the spectral efficiency. While for WSMS architecture in~\cite{8356240}, $d_s$ is directly  approximated by $\sqrt{\lambda D/2}$, which does not maximize the spectral efficiency. Second, in this work, the hybrid beamforming matrices $\textbf{P}_{\rm A}$, $\textbf{P}_{\rm D}$, $\textbf{C}_{\rm A}$, and $\textbf{C}_{\rm D}$ are designed by the proposed optimal closed-form solution, which performs better than the non-optimal hybrid beamforming algorithm in~\cite{8356240}.

\subsection{Performance Evaluation of the Proposed Algorithms}
Then, we analyze the performance of the proposed algorithms in this work. First is the optimal closed-form solution in Sec.~\ref{section_beamforming}, which designs the hybrid beamforming matrices $\textbf{P}_{\rm A}$, $\textbf{P}_{\rm D}$, $\textbf{C}_{\rm A}$, and $\textbf{C}_{\rm D}$ to maximize the spectral efficiency. Second is the DLR algorithm in Sec.~\ref{section_design_subarray}, which determines the number of subarrays $k$ and the subarray spacing $d_s$ to maximize the spectral efficiency. 

\begin{figure}
	\centering
	\begin{tabular}{cc}
		\begin{minipage}[t]{0.449\linewidth}
			\includegraphics[width = 1\linewidth]{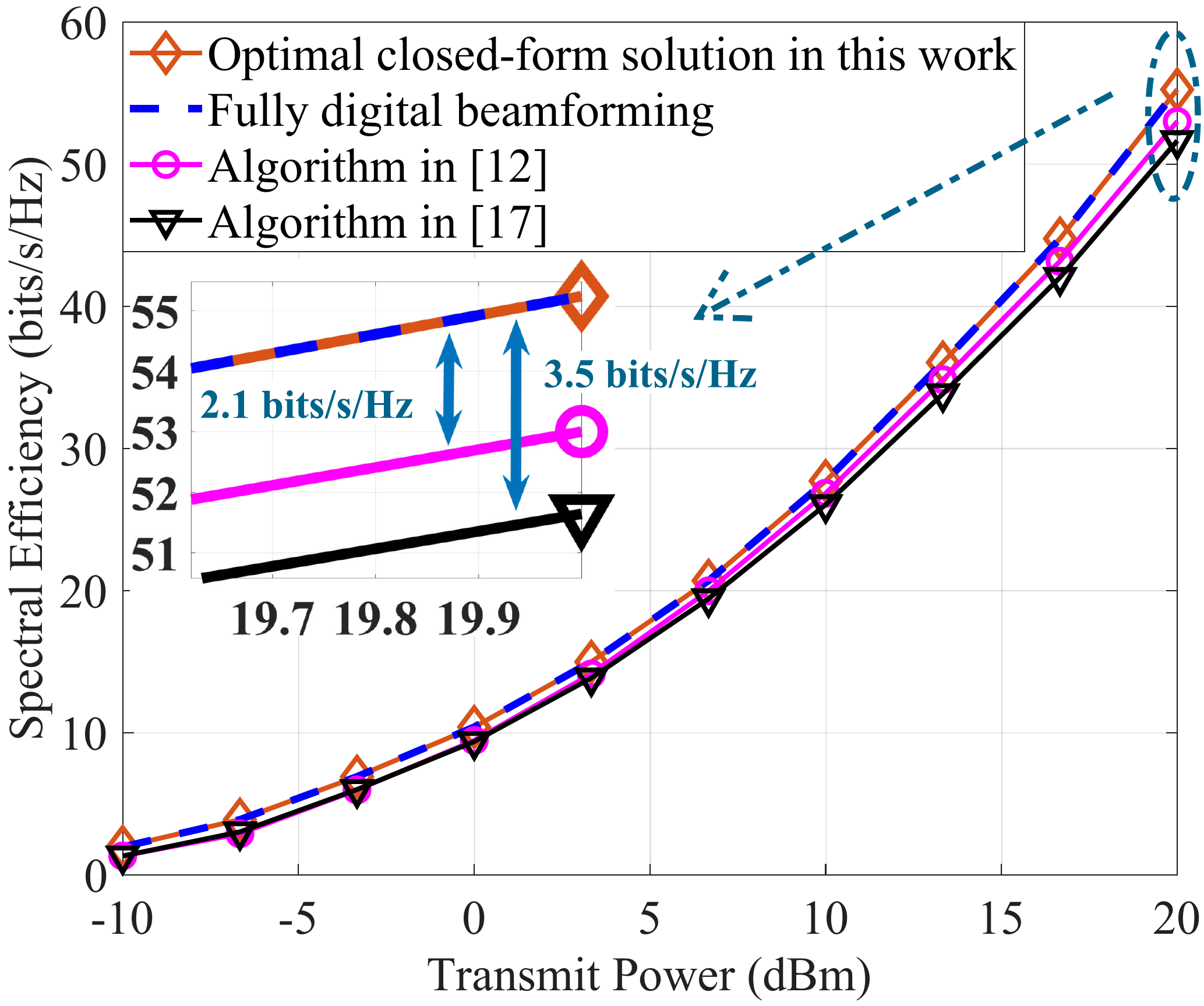}
			\captionsetup{font={footnotesize}}
			\caption{Spectral efficiency of THz WSMS architecture with different hybrid beamforming algorithms, $N_t=N_r=1024$, $D=60$m.}
			\label{fig_SE_vs_HBF}
		\end{minipage}
		\begin{minipage}[t]{0.45\linewidth}
			\includegraphics[width = 1\linewidth]{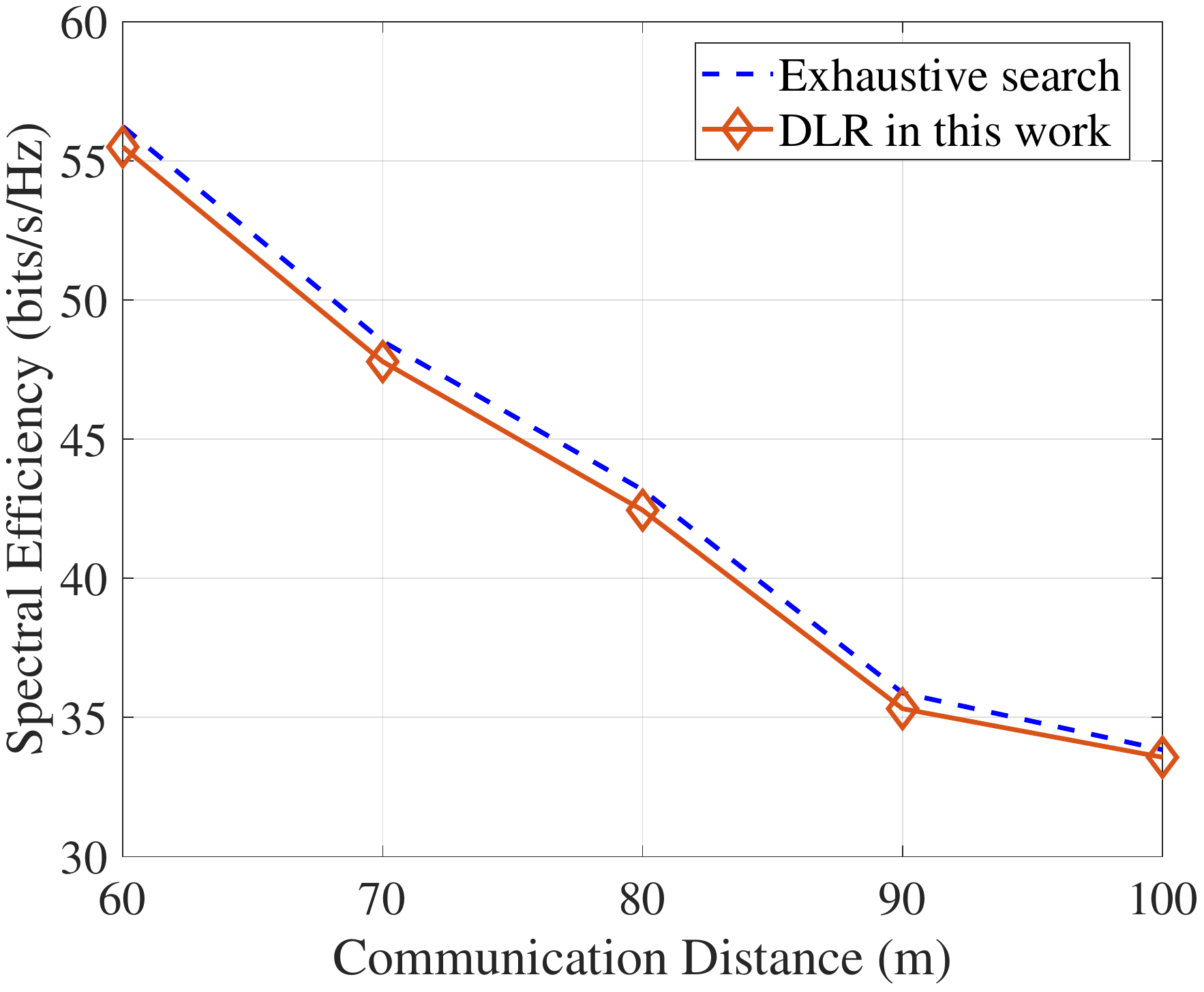}
			\captionsetup{font={footnotesize}}
			\caption{Spectral efficiency of THz WSMS architecture with DLR and exhaustive search algorithms, $N_t=N_r=1024$, $\rho=$ 20 dBm.}
			\label{fig_SE_vs_DLR}
		\end{minipage}
	\end{tabular}
	\vspace{-9.5mm}
\end{figure}

Fig.~\ref{fig_SE_vs_HBF} shows the performance of the proposed optimal closed-form solution. To fairly evaluate the performance of hybrid beamforming algorithms, all algorithms are implemented in the same WSMS architecture, where $k$ and $d_s$ are calculated by DLR algorithm. It is well known that the spectral efficiency of the fully digital beamforming is the upper bound of all hybrid beamforming algorithms. Our proposed optimal closed-form solution can achieve the same spectral efficiency with the fully digital beamforming, which demonstrates that the proposed optimal closed-form solution is indeed optimal. Compared to the algorithm in~\cite{8356240}, the spectral efficiency of the optimal closed-form solution is higher, e.g., 3.5 bits/s/Hz higher when $\rho=20$ dBm. It is worth noting that the analog beamforming matrices $\textbf{P}_{\rm A}$ and $\textbf{C}_{\rm A}$ have a block-diagonal structure as shown in~\eqref{structure of PA DAoSA}. Hence, most of the existing hybrid beamforming algorithms proposed for FC and AoSA architectures can not be directly applied to WSMS architecture. Therefore, we select the algorithm in~\cite{7389996}, which calculates $\textbf{P}_{\rm A}$ and $\textbf{C}_{\rm A}$ element-by-element and can overcome the block-diagonal constraint, as a comparison. Compared to the algorithm in~\cite{7389996}, our optimal closed-form solution achieves a higher spectral efficiency, e.g., 2.1 bits/s/Hz when $\rho=20$ dBm.

Moreover, the computational complexity of the proposed optimal closed-form solution is $\mathcal{O}((N_t+N_r)N_s^2)$. By contrast, the computational complexities of the algorithm in~\cite{8356240} and algorithm in~\cite{7389996} are $\mathcal{O}((N_t^3+N_r^3)N_s)$ and $\mathcal{O}(N_t^3+N_r^3)$, respectively.
In THz band, the number of antennas $N_t$ and $N_r$ are usually much larger than the number of data streams $N_s$. Hence, the computational complexity of our proposed optimal closed-form solution is substantially lower than the algorithms in~\cite{8356240} and~\cite{7389996}.

Fig.~\ref{fig_SE_vs_DLR} evaluates the performance of the proposed DLR algorithm. The spectral efficiency of y-axis is the objective function of subproblem P2 in~\eqref{P1}. Since the subproblem P2 is firstly formulated in this work, there is no existing algorithms proposed for P2. Therefore, we select the exhaustive search algorithm as the benchmark for comparison.
For all communication distances, the DLR algorithm can achieve similar spectral efficiency with the optimal exhaustive search algorithm. Therefore, without loss of generality, the DLR algorithm can design $k$ and $d_s$ to maximize the spectral efficiency near-optimally. The computational complexity of the DLR algorithm is much lower than the exhaustive search algorithm, since the DLR algorithm transfers the subproblem P2 into a more tractable form~\eqref{C_LOS_4} and carries the gradient descend method, rather than exhaustively searching. 

\subsection{Energy Efficiency Comparison}
Next, we evaluate the energy efficiency of the THz WSMS architecture with the proposed algorithms in this work. The energy efficiency is defined as the ratio between spectral efficiency and power consumption. The power consumption of the hardware components in THz WSMS architecture at the transmitter can be expressed as
\begin{equation}
	\rm P_{Tx}=P_{PA}N_{PA}+P_{PC}N_{PC}+P_{PS}N_{PS}+P_{RF}N_{RF}+P_{DAC}N_{DAC}+P_{BB}N_{BB},
	\label{calculation_power}
\end{equation}
where the parameters are provided in TABLE~\ref{power_device}. The calculation of power consumption of the other architectures also follows~\eqref{calculation_power} while with some justifications as follows. For FC architecture, ${\rm N_{PS}}=N_tL_t$. For AoSA architecture, ${\rm N_{PS}}=N_t$ and $\rm N_{PC}=0$. For LoS MIMO architecture which is fully-digital, $\rm N_{PS}=N_{PC}=0$ and ${\rm N_{DAC}=N_{RF}}=N_t$. The value of $L_t$ for different architectures is provided in the second paragraph of Sec.~\ref{section_simulation}.
The power consumption $\rm P_{Rx}$ at the receiver side can be calculated similarly by using the parameters of the hardware components at the receiver~\cite{DAoSA_JSAC_2020,8733134}. Therefore, the total power consumption of the system is ${\rm P_{Tx}+P_{Rx}}+\rho$.

\begin{table}
	\centering
	\captionsetup{font={footnotesize}}
	\caption{Device power and the quantity of the devices used in THz WSMS architecture at the transmitter~\cite{DAoSA_JSAC_2020,8733134}.}
	\begin{tabular}{|c|p{200pt}<{\centering}|c|}
		\hline
		Device&Individual device power [mW], around 0.3 THz&Quantity\\
		\hline Power amplifier &${\rm P}_{\rm PA}=40$& ${\rm N}_{\rm PA}=N_t$\\
		\hline Power combiner&${\rm P}_{\rm PC}=6.6$ & ${\rm N}_{\rm PC}=N_t$\\	
		\hline Phase shifter &${\rm P}_{\rm PS}=42$& ${\rm N}_{\rm PS}=N_tL_t/k$\\
		\hline RF chain&${\rm P}_{\rm RF}=26$& ${\rm N}_{\rm RF}=L_t$\\
		\hline DAC&${\rm P}_{\rm DAC}=110$& ${\rm N}_{\rm DAC}=L_t$\\
		\hline Baseband&${\rm P}_{\rm BB}=200$& ${\rm N}_{\rm BB}=1$\\  
		\hline
	\end{tabular}
	\label{power_device}
	\vspace{-7.5mm}
\end{table}
\begin{figure}
	\centering
	\captionsetup{font={footnotesize}}
	\includegraphics[scale=0.45]{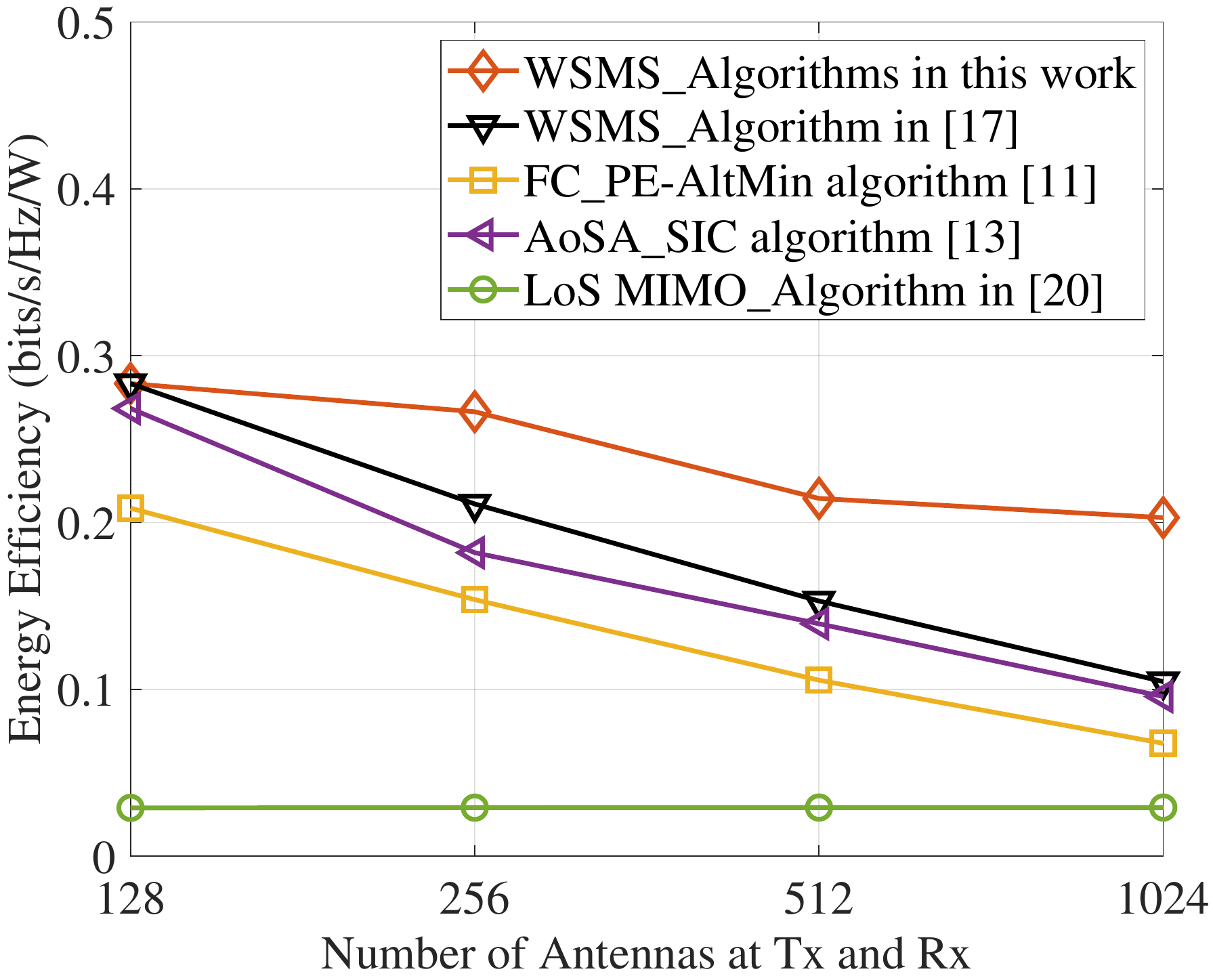}
	\caption{Energy efficiency versus number of antennas. $N_t=N_r$, $D=60$m, $\rho=20$ dBm.}
	\label{fig_EE_NtNr}
	\vspace{-7.5mm}
\end{figure}
Fig.~\ref{fig_EE_NtNr} evaluates the energy efficiency of THz WSMS architecture with different number of antennas. The power consumption of all architectures grows linearly with the number of antennas, as shown in~\eqref{calculation_power}. While the relationship between the spectral efficiency~\eqref{R} and number of antennas has a logarithmic function. Therefore, the energy efficiency of the THz WSMS, FC, and AoSA architectures decreases with the growth of the number of antennas. For LoS MIMO architecture, owing to the poor spectral efficiency, the energy efficiency is very low for all number of antennas. Particularly, since the spectral efficiency of the THz WSMS architecture with the proposed algorithms in this work is significantly higher than the other architectures, its energy efficiency also outperforms the other architectures substantially. For instance, when $N_t=N_r=1024$, the energy efficiency of the THz WSMS architecture with the proposed algorithms in this work is at least 1.91 times of those of the other architectures.

\section{Conclusion}
\label{section_conclusion}
In this paper, a THz WSMS hybrid beamforming architecture is investigated to joint utilize the inter-path multiplexing and intra-path multiplexing.  We first prove that the joint inter-path and intra-path multiplexing gain is $kN_p$, which is $k$ times of the multiplexing gain of the existing hybrid beamforming architectures, where $k$ and $N_p$ denote the number of widely-spaced subarrays in WSMS and number of multipath of the channel. Next, we elaborate that both $k$ and the subarray spacing $d_s$ contribute to the spectral efficiency of the WSMS architecture significantly. Inspired by this, we formulate a novel design problem as designing $k$, $d_s$, and the hybrid beamforming matrices to maximize the spectral efficiency. The design problem is further decomposed into a hybrid beamforming subproblem P1 and an array configuration subproblem P2. Through exploiting the structure of channel, we propose an optimal closed-form solution for hybrid beamforming problem P1. Moreover, we propose a DLR algorithm for array configuration problem P2, by utilizing the dominant LoS property of the THz band.

Extensive simulation results show that, the spectral efficiency of the THz WSMS architecture in this work is 189\% higher than the existing FC and AoSA architectures, thanks to the $k$ times of the multiplexing gain provided by the joint utilization of inter-path and intra-path multiplexing. Moreover, the spectral efficiency of THz WSMS architecture in this work is 96\% higher than the existing special case WSMS architecture with $k=2$, due to the careful design of $k$ and $d_s$ through the proposed DLR algorithm. The energy efficiency of THz WSMS architecture in this work is 91\% higher than counterpart architectures. Furthermore, we demonstrate that the performances of the proposed optimal closed-form solution and the DLR algorithm are optimal and near-optimal, respectively. Both of these two algorithms have much lower computational complexities than the existing algorithms.

\bibliographystyle{IEEEtran}
\bibliography{IEEEabrv,references}
\end{document}